\documentclass[aps,prx,twocolumn,superscriptaddress,groupedaddress,nofootinbib]{revtex4}
\usepackage{graphicx}  
\usepackage{dcolumn}   
\usepackage{bm}        
\usepackage{amssymb}   
\usepackage{amsmath}
\usepackage{dsfont}
\usepackage{xcolor}
\usepackage{amsthm}
\usepackage{enumitem}
\usepackage{tensor}
\usepackage[utf8]{inputenc}
\usepackage{float}
\usepackage{graphicx}
\usepackage{dcolumn}
\usepackage{bm}

\usepackage[colorlinks]{hyperref}
\hypersetup{
     breaklinks=true,
    pdfstartview={FitH},    
    colorlinks=true,       
    linkcolor=blue,          
    citecolor=red,        
    filecolor=magenta,      
    urlcolor=blue,           
    anchorcolor=green,      
    linktocpage=true
}

\newcommand{\be}{\begin{equation}}\newcommand{\ee}{\end{equation}}
\newcommand{\bea}{\begin{eqnarray}}\newcommand{\eea}{\end{eqnarray}}
\newcommand{\brr}{\begin{array}}\newcommand{\err}{\end{array}}
\newcommand{\bit}{\begin{itemize}}\newcommand{\eit}{\end{itemize}}
\newcommand{\ben}{\begin{enumerate}}\newcommand{\een}{\end{enumerate}}

\newcommand{\ba}{\begin{array}}
\newcommand{\ea}{\end{array}}

\definecolor{darkred}{rgb}{.8,0,0}

\definecolor{darkblue}{rgb}{0,0,0.7}

\def\1{{_{1}}}\def\2{{_{2}}}

\def\noHe0{:\;\!\!\;\!\!:H_e(0):\;\!\!\;\!\!:}
\def\noHm0{:\;\!\!\;\!\!:H_\mu(0):\;\!\!\;\!\!:}

\begin{document}


\title{$q$-Exponential Random Graphs: higher-order networks from simple constraints}

\author{David Dob\'{a}\v{s}}
 \email{dobasdav@fjfi.cvut.cz}
\affiliation{%
 FNSPE,
Czech Technical University in Prague, B\v{r}ehov\'{a} 7, 115 19, Prague, Czech Republic
}%
\author{Diego  Garlaschelli}
\email{garlaschelli@lorentz.leidenuniv.nl}
\affiliation{
Lorentz Institute for Theoretical Physics, University of Leiden
Niels Bohrweg 2, 2333 CA Leiden, The Netherlands}
\affiliation{IMT School for Advanced Studies
Piazza S. Francesco 19, 55100 Lucca, Italy
}
\author{Petr Jizba}%
 \email{p.jizba@fjfi.cvut.cz}
\affiliation{%
 FNSPE,
Czech Technical University in Prague, B\v{r}ehov\'{a} 7, 115 19, Prague, Czech Republic
}%



\date{May 29, 2026}

\begin{abstract}
Exponential Random Graphs (ERGs) are among the most widely used network models, derived as principled least-bias graph ensembles that maximize Shannon entropy under constraints on the expected values of given structural properties. 
However, it has been recently (re)discovered that, in absence of additional information privileging Shannon entropy, the most agnostic inferential construction should maximize the broader class of Uffink entropies. The resulting entropy-maximizing distribution changes from the exponential (Boltzmann--Gibbs) to the so-called $q$-exponential one.
Since maximizing Shannon entropy may produce an unjustified independence between degrees of freedom, here we investigate how the most popular ERGs with independent edges (namely, the Erd\H{o}s--R\'enyi and configuration models) generalize to higher-order $q$-Exponential Random Graphs with dependent edges in the non-Shannon case, while keeping their defining constraints (number of links and degree sequence respectively) unchanged.
We find features, such as a phase transition between sparse and dense regimes, that are absent in the original ERGs but typical of higher-order networks, plus novel phenomena such as richer assortativity and clustering profiles, which allow for the coexistence of link sparsity and triadic closure.
These results show that higher-order networks do not necessarily require higher-order constraints, as they naturally arise from simpler ones in a framework that is even more agnostic than Shannon’s.
\end{abstract}

\maketitle


%

\section{Introduction\label{sec:intro}}
%

Exponential Random Graphs (ERGs) were initially proposed for social network analysis~\cite{Holland1981,strauss1986general,wasserman1994social,anderson1999p,robins2007introduction} and have been later re-derived~\cite{Park2004} as principled unbiased models of graphs that can be successfully used, among other applications, for network reconstruction, pattern detection and combinatorial enumeration~\cite{bianconi2009entropy,fronczak2012exponential,Squartini2017,coolen2017generating,Cimini2019}.
ERGs are defined as canonical ensembles of graphs that maximize Shannon entropy~\cite{Shannon1948} under constraints on the expected value of (in principle) any desired network property.
Being a family of models based on the choice of the defining constraints, ERGs are very flexible and in fact ubiquitous in network science: prominent examples include the Erd\H{o}s--R\'enyi random graph model~\cite{Erdoes1959,erdHos1960evolution} which can be obtained from a single constraint on the expected density of links in the graph, the Configuration Model~\cite{Park2003} which corresponds to multiple constraints on the expected degrees of all nodes separately (i.e., on the expected degree sequence), and the (Degree-Corrected~\cite{fronczak2013exponential,karrer2011stochastic,peixoto2019bayesian}) Stochastic Block Model~\cite{wasserman1994social,anderson1992building} which separately constrains (the degree sequence in addition to) the densities of links within and across different blocks of nodes, thus producing graphs with community structure.\\

ERGs permeate network science because, depending on the chosen constraints and how they are enforced, they can be used as unbiased null models to test for higher-order properties (such as reciprocity~\cite{Holland1981,garlaschelli2004patterns,squartini2013reciprocity,gallo2025patterns}, triadic closure~\cite{Newman2003,fronczak2007phase}, network motifs~\cite{milo2002network}, core-periphery structure~\cite{holme2005core}, communities~\cite{newman2006modularity}, or other statistically validated patterns~\cite{saracco2017inferring,gallo2025statistically}), as structural models to replicate given features and identify structure via optimal parameter fitting~\cite{Holland1981,strauss1986general,hunter2008ergm,coolen2017generating,wasserman1994social,Squartini2011,peixoto2019bayesian}, and finally as instruments for network reconstruction~\cite{Squartini2018,cimini2021reconstructing}, link imputation~\cite{krause2018missing}, and other inferential tasks~\cite{Squartini2017}.

\subsection{The maximum-entropy  foundation of ERGs\label{sec:mep}}
The reason why ERGs are so widely accepted in network science comes from their rigorous derivation from the Maximum Entropy Principle (MEP), originally introduced heuristically by Jaynes~\cite{Jaynes1957,Jaynes1957b} and later axiomatically formalized by Shore and Johnson~\cite{Shore1980,Shore1981}. The MEP states that, given partial (aggregate) empirically available information about a (large) system --- phrased in terms of expectation values (often low-order correlations or cumulants) ---  the least biased inference about the underlying microscopic state is a necessarily probabilistic. It is obtained by selecting the probability distribution, over all admissible microscopic states, that maximizes Shannon entropy, subject to the imposed constraints encoding the available information. 
In Jaynes' original formulation of the MEP, other types of constraints, such as
escort means, quasi-linear means or non-inductive prior
information (e.g., constraints involving the Lipschitz--H\"{o}lder regularity of probability distributions or the Hausdorff dimension of the underlying state space) are not considered. 
As well known~\cite{Jaynes1957,Park2004,Cimini2019} (and briefly summarized below), the specific form of Shannon entropy implies that the resulting entropy-maximizing probability has an exponential dependence on the chosen constraints, giving (in the context of network models) ERGs their name.\\

But why is the MEP a rigorous construction in the first place? In turn, this comes from certain guarantees enforced via the axiomatic definition of Shannon entropy~\cite{Khinchin1957,Shannon1948} as the amount of uncertainty encoded in a probability distribution. In the language of random graphs relevant to this paper, if ${\mathcal{G}_{n}}$ denotes a set of graphs of interest (e.g., undirected/directed, binary/weighted, etc.) on $n$ nodes and $P\equiv\{P(G)\}_{G\in\mathcal{G}_{n}}$ denotes a discrete probability distribution attaching a probability $P(G)$ to each graph $G\in{\mathcal{G}_{n}}$, then Shannon entropy is the only scalar function $S[P]$ of the elements of $P$ that respects the following four Shannon--Khinchin (\emph{SK}) axioms~\cite{Khinchin1957}:
\begin{itemize}
\item \textit{SK1 (continuity):} $S[P]$ is continuous in the entries of $P$. 
\item \textit{SK2 (maximality):} $S[P]$ is maximal when $P$ is the uniform distribution over ${\mathcal{G}_{n}}$. 
\item \textit{SK3 (expansibility):} $S[P]$ does not change if an extra microscopic configuration $G'$ with zero probability $P(G')=0$ is added to the support ${\mathcal{G}_{n}}$.
\item \textit{SK4 (additive separability):} if $P(G)$ is interpreted as a joint distribution of two random variables (here obtained by partitioning the random graph $G$ into two non-overlapping subsets $G_1$ and $G_2$ of pairs of nodes, with $G=G_1\cup G_2$ and $G_1\cap G_2=\emptyset$) with marginal distributions $Q(G_1)$ and $R(G_2)$ respectively, then $S[P]$ separates additively as 
\begin{equation}
S[P] \ = \ S[Q] \ + \ S[R|Q]\, ,
\label{additive}
\end{equation}
where $S[R|Q]$ is the conditional entropy defined as the mean, over $Q(G_1)$, of the entropy of the conditional distribution of $G_2$, given $G_1$.
Note that in particular, if $G_1$ and $G_2$ are independent, i.e. $P(G)= Q(G_1)R(G_2)$, then $S[P]$ is simply the sum of $S[Q]$ and $S[R]$, i.e.
\begin{equation}
P(G)\ = \ Q(G_1)R(G_2)\;
\Rightarrow \; S[P] \ = \  S[Q] \ + \ S[R]\, .
\label{additive0}
\end{equation}
\end{itemize}
It is possible to show~\cite{Khinchin1957} that, up to an inessential positive prefactor, the only quantity realizing the above four \emph{SK} axioms is Shannon entropy 
\begin{equation}
    S[P]\ = \ -\sum_{G \in \mathcal{G}_{n}} P(G)\ln P(G)\, ,
    \label{S1}
\end{equation}
which represents a cornerstone in information theory~\cite{Cover2006} and statistical physics, where it coincides with Gibbs entropic functional~\cite{gibbs2014elementary} and (upto a sign) with Boltzmann's H-function~\cite{Boltzmann1872}.\\

With the above properties of Shannon entropy in mind, the rationale behind the MEP is that, in presence of only aggregate information about the state of a system, the least biased probability distribution characterizing the unobserved microscopic states should be the one that encodes maximal uncertainty, besides reproducing (in expectation) the observable aggregate properties over the distribution itself~\cite{Jaynes1957}.
This is why ERGs give a precise meaning to otherwise vague definitions such as `random graphs with given link density', `random graphs with given node degrees', or `random graphs with given community structure': all such  models can be rigorously understood as the least biased, maximum-entropy graph ensembles with given expected properties~\cite{Park2004,bianconi2009entropy,Squartini2017}. The Erd\H{o}s--R\'enyi model~\cite{Erdoes1959,erdHos1960evolution}, the (canonical) Configuration Model~\cite{Park2003} and the Stochastic Block Model~\cite{fronczak2013exponential,karrer2011stochastic,peixoto2019bayesian} represent precisely such ensembles, respectively.

\subsection{A generalized maximum entropy principle\label{sec:qmep}}
In this paper, we introduce an additional element to the definition of ERGs, which descends from the recent fascinating (re)discovery that the MEP admits a more agnostic construction, based in turn on a more general family of entropy functionals with desirable inferential properties~\cite{Uffink1995,Jizba2019,Jizba2020}. Indeed, while Shannon entropy is the only entropy respecting the four \emph{SK} axioms as stated above, there is an important generalization that  arises from relaxing the  additivity requirement expressed in Eq.~\eqref{additive} of \emph{SK4}.
It should be noted at this point that such generalization of \emph{SK4} has been a matter of discussion in the statistical physics literature for decades (see, e.g.~\cite{Thurner} and references therein). This is mainly because it has been often regarded as the relaxation of Eq.~\eqref{additive0} rather than of Eq.~\eqref{additive} --- hence as a transition to an unavoidably non-extensive (or rather non-additive) thermodynamics~\cite{Tsallis2009}, such as the one based on Tsallis entropy~\cite{Tsallis1988}
\begin{equation}
T_q[P] \ \equiv \  \frac{1}{1-q} \bigg(\sum_{G \in \mathcal{G}_{n}}  P^q(G)-1\bigg)\, ,
\label{tsallis}
\end{equation} 
which indeed violates Eq.~\eqref{additive0}. 
However, while Eq.~\eqref{S1} implies Eq.~\eqref{additive} which in turn implies Eq.~\eqref{additive0}, there exist generalizations of Eq.~\eqref{S1} that violate Eq.~\eqref{additive}, while still realizing Eq.~\eqref{additive0}, hence remaining completely additive over independent systems. The most notable such extension is R\'enyi entropy~\cite{renyi1961measures,Renyi:76}, defined as
\begin{equation}
R_q[P] \ \equiv \  \frac{1}{1-q} \ln \sum_{G \in \mathcal{G}_{n}}  P^q(G)\, .
\label{renyi}
\end{equation}

To explore possible generalizations of Eq.~\eqref{S1} systematically, it was shown in Refs.~\cite{Jizba2019,Jizba2020} that if the sum in Eq.~\eqref{additive} of \emph{SK4} is replaced with the following so-called Kolmogorov--Nagumo operation\footnote{Given a set $M$ and a bijection $f^{-1}:M\mapsto N\subset \mathbb{R}$, the generalized Kolmogorov-Nagumo arithmetics is defined as follows:
\begin{eqnarray*}
    x \oplus_f y &=& f(f^{-1}(x)+f^{-1}(y)), \\
    x \ominus_f y &=& f(f^{-1}(x)-f^{-1}(y)), \\
    x \otimes_f y &=& f(f^{-1}(x)f^{-1}(y)), \\
    x \oslash_f y &=& f(f^{-1}(x)/f^{-1}(y)).
\end{eqnarray*}} 
\begin{equation}
S[P] \ = \ S[Q]\otimes_f S[R|Q]\, ,
\label{additivef}
\end{equation}
(arising in the context of generalized arithmetics~\cite{kolmogorov1930notion,nagumo1930klasse}), then the most general entropy consistent with the four \emph{SK} axioms (with modified \emph{SK4}) becomes 
 \begin{eqnarray}
&&S^{(f)}_q[P] \ = \ f(U_q[P])\, ,\label{uffink_entropy}\\[2mm]
&&U_q[P] \ = \ \bigg(\sum_{G \in \mathcal{G}_{n}} P^q(G)\bigg)^\frac{1}{1-q}\, ,
\label{uffink}
\end{eqnarray}
where $f$ is an arbitrary strictly increasing and positively supported function, $U_q[P]$ is what is called  (for reasons that will be clear in a moment) the Uffink functional~\cite{Uffink1995}, and $q>0$ is what we may call an `entropic parameter'~\cite{somazzi2025learn}. 
Indeed, the choice $f(x)=\ln x$ retrieves R\'enyi entropy in Eq.~\eqref{renyi}:
\begin{equation}
S^{(\ln)}_q[P] \ = \ R_q[P]\, .
\end{equation} 
Similarly, the choice $f(x)= \ln_q (x)$, where
\begin{equation}
\ln_q(x) \ \equiv \ \frac{x^{1-q}-1}{1-q}\, ,
\label{qlog}
\end{equation}
is the so-called ``$q$-logarithm'' (not to be confused with the ordinary logarithm of $x$ to base $q$), retrieves Tsallis entropy in Eq.~\eqref{tsallis}:
\begin{equation} 
S^{(\ln_q)}_q[P] \ = \ T_q[P]\, .
\end{equation} 
For both entropies, the limit $q \to 1$ recovers the Shannon entropy given in Eq.~\eqref{S1}, i.e.
\begin{eqnarray}
\lim_{q\to 1} S^{(\ln)}_q[P]\ = \ \lim_{q\to 1} S^{(\ln_q)}_q[P] \ = \ S[P]\, .
\label{wShannon}
\end{eqnarray}

The family of entropies given by Eq.~\eqref{uffink} is quite general (see~\cite{Jizba2019,Jizba2020,Jizba2019b,somazzi2025learn}). 
One might nonetheless wonder whether the modified form of axiom \emph{SK4} in Eq.~\eqref{additivef}, and hence the resulting modified entropy in Eq.~\eqref{uffink}, is merely an arbitrary construction or instead arises from a deeper mathematical consistency requirement.
Crucially, it turns out that this modification is in fact necessary in order to recover the most general form of entropy consistent with another set of axioms, introduced by Shore and Johnson (\emph{SJ})~\cite{Shore1980,Shore1981}, which ensure the logical consistency of the MEP within Statistical Inference Theory (SIT). At this point, it is important to emphasize that, in SIT, entropy functionals serve primarily as technical tools for the unbiased assignment of probability distributions consistent with a given set of constraints. Indeed, one may argue that the maximum-entropy distribution is the primary object in SIT, while the entropy itself plays a secondary role, without an independent operational interpretation within the framework. This stands in contrast to information theory, where entropy is a primary quantity, often endowed with a clear operational meaning, for example in terms of coding theorems. Even if entropy play a secondary role in SIT, it must still satisfy a minimal set consistency conditions. Specifically, Shore and Johnson formulated their axioms not as requirements on the mathematical form of the entropy itself, but rather as conditions on the properties of the probability distribution obtained by maximizing the entropy under given constraints~\cite{Shore1980,Shore1981}:
\begin{itemize}
\item \textit{SJ1 (uniqueness):} The result (i.e., the entropy-maximizing  distribution) should be unique.
\item \textit{SJ2 (invariance):} It shouldn't matter in which coordinate system one
accounts for new information.
\item \textit{SJ3 (Subset independence):} It should not matter whether one treats disjoint subsets of system states in terms of separate conditional distributions or in terms of the full distribution.
\item \textit{SJ4 (System independence):} It should not matter whether one accounts for independent information (i.e., constraints) about independent systems separately in terms of marginal distributions or together in terms of a single joint distribution.
\item \textit{SJ5 (Maximality):} In absence of any prior information, the uniform distribution should be the solution.\footnote{Actually, Shore and Johnson defined the maximality axiom only implicitly. Indeed, starting from the principle of minimum cross-entropy, they introduced the MEP as its equivalent in the case where the prior distribution is uniform. For this reason, even if not explicitly axiomatized, they considered the posterior $P$ to be equal to the uniform distribution (i.e. the same as the prior) when no information is available.}
\end{itemize}

Curiously, Shore and Johnson incorrectly claimed that Shannon entropy in Eq.~\eqref{S1} is the only choice compatible with their axioms~\cite{Shore1980}, a statement that has for a long time suggested the equivalence of the \emph{SJ} axioms and the original \emph{SK} ones. 
However, Uffink~\cite{Uffink1995} later identified an error behind Shore and Johnson's initial conclusion, which was due to an unjustified hidden assumption: when considering the availability of distinct pieces of information about two subsystems of the same system, Shore and Johnson incorrectly implied that the resulting maximum-entropy joint probability factorises, \emph{de facto} applying \emph{SJ4} even when the independence of the two systems is not guaranteed~\cite{Uffink1995,Jizba2019}. 
However, having disjoint pieces of information about two subsystems obviously does not guarantee that the subsystems are independent.
Unfortunately, as we discuss in detail in this paper, this is precisely what happens with the exponential probability distributions maximizing Shannon entropy, and therefore with the whole class of ERGs used so far in network science.\\

Remarkably, Uffink~\cite{Uffink1995} showed that, if the hidden unjustified assumption of independence is dropped, then the entropy resulting from the \emph{SJ} axioms is not necessarily the Shannon entropy, but rather the broader class of entropies given in Eq.~\eqref{uffink} --- justifying the choice of the name ``Uffink functional'' for the quantity $U_q[P]$.
Consequently, the \emph{SJ} axioms are, in fact, equivalent to the \emph{modified} \emph{SK} axioms [as defined via Eq.~\eqref{additivef}]~\cite{Jizba2019,Jizba2020}, rather than the original ones. Both sets of axioms therefore characterize the full class of Uffink-type entropies as inferentially admissible, rather than singling out the Shannon entropy as the unique solution.
Shannon entropy is the only unique solution when, in addition to the original \emph{SJ} axioms, an additional assumption of independence is made. This also means that, in the equivalent language of \emph{SK} axioms, the additional independence assumption corresponds to the restriction from the general Kolmogorov-Nagumo separability in Eq.~\eqref{additivef} to the additive separability in Eq.~\eqref{additive}, i.e. to a specific choice of $f$ with $q\to 1$ in the Uffink functional $S^{(f)}_q[P]$ defined in Eq.~(\ref{uffink_entropy}). 
For this reason, $q$ can be interpreted as a parameter controlling the degree of dependency between otherwise independent degrees of freedom in the Shannon ($q=1$) case: values $q\ne 1$ can induce positive or negative correlations among otherwise uncorrelated variables.
Similarly, as we will show in more detail, the Kolmogorov--Nagumo function $f$ determines the scaling behavior of the entropy for uniform distributions~\cite{Jizba2020}.

\subsection{From exponential to $q$-exponential ensembles\label{sec:qexp}}

Accepting the extended family of entropies in Eq.~\eqref{uffink} as the most general inferential solution implies acknowledging that the MEP should be correspondingly generalized to the search for the most general probability distribution that, given a set of input information, maximizes $S^{(f)}_q[P]$, rather than being restricted to the Shannon entropy $S[P]$~\cite{Uffink1995,Jizba2019,Jizba2020,somazzi2025learn,morales2021generalization}.
In other words, in the absence of additional information that would privilege the Shannon form, a maximally agnostic inference procedure should instead consider the broader class of Uffink-type entropies. This is because maximizing the Shannon entropy may implicitly enforce independence between degrees of freedom, which need not be justified in general.
Clearly, the resulting maximum-entropy solutions therefore constitute a family of probability distributions that includes, but is not limited to, the standard exponential distributions obtained from Shannon entropy.

Note that, conveniently, the resulting generalized probability distribution is independent of $f$ --- since $f$ is assumed to be monotonically increasing, the maximizer of $S^{(f)}_q[P]$ is unaffected by the specific choice of $f$ --- but it does depend on $q$. We are therefore dealing with a relatively simple, one-parameter extension of the standard exponential distribution.
The ensuing generalization is known as $q$-exponential distribution (which indeed maximizes Tsallis entropy $T_q[P]$, R\'enyi entropy $R_q[R]$ and Uffink entropy $U_q[R]$ altogether~\cite{Tsallis1988,Tsallis2009,Uffink1995,Jizba2019,Jizba2020,somazzi2025learn,morales2021generalization}). The $q$-exponential function is defined as the inverse of the $q$-logarithm $\ln_q (x)$ defined in Eq.~\eqref{qlog}, i.e.
\begin{equation}
\exp_q(x) \ = \  [1 \ + \  (1 - q)\ \! x]^{1/(1-q)}_+, \quad q\in \mathbb{R}\, ,
\label{qexp}
\end{equation}
(with $[z]_+\equiv \max\{z,0\}$ guaranteeing that the probability distribution is single-valued and real).
In the limit $q\to 1$, the ordinary exponential function is retrieved:
\begin{equation}
\lim_{q\to 1}\exp_q(x) \ = \ \exp(x)\, ,
\end{equation}
consistently with the fact that, in the same limit, Shannon entropy is retrieved as previously noticed in Eq.~\eqref{wShannon}.\\

Based on the above considerations, in this paper we make the natural logical step of introducing the $q$-exponential generalization of ERGs, which we denote as $q$-Exponential Random Graphs ($q$-ERGs), as the most general and agnostic class of maximum-entropy network models with given constraints. 
Besides the general theoretical and applied relevance of exploring $q$-ERGs, an aspect of particular interest is how the independence of links in Shannonian ERGs gets replaced by nontrivial dependencies in $q$-ERGs.
This is particularly relevant in light of the recent emphasis on higher-order networks and the several attempts of introducing parsimonious models capturing them in terms, e.g. of (random) simplicial complexes and hypergraphs~\cite{battiston2022higher,bick2023higher,bianconi2021higher,boccaletti2023structure,peixoto2026graphs}.

The layout of the present paper is as follows. In Sec.~\ref{sec:shannon},  we briefly revisit some fundamentals of the MEP applied to graphs, leading to Shannonian ERG models. In particular we review how the popular Erd\H{o}s--R\'enyi models and the Configuration Model emerge in this context as the most important models with independent edges. 
In Sec.~\ref{sec:qERGs} we introduce our generalized family of $q$-ERGs which is obtained when the MEP is extended to the Uffink entropy. 
In particular, we explicitly derive the generalized $q$-exponential graph probability distribution and introduce various technical tools and concepts that become necessary in the non-Shannon ($q\ne1$) case, including the concept of effective Hamiltonian and a reparametrization technique that eliminates self-referentiality.
In Sec.~\ref{sec:qER}, we apply the $q$-ERG framework to obtain the natural generalization of the Erd\H{o}s--R\'enyi model, analyzing its large-$N$ behavior, the emergence of edge dependencies that are typical of models with higher-order constraints, and the accompanying phase transition. 
In Sec.~\ref{sec:tsallis_park_newman} we carry out the corresponding $q$-exponential generalization of the Configuration Model, identifying novel results such as the possibility of enhancing the local clustering coefficients and altering the assortativity properties by varying $q$, while keeping the degree sequence constrained.
We finally offer our conclusions in Sec.~\ref{sec:conclusions}.

\section{Traditional Shannonian ERGs\label{sec:shannon}}
%
Here we briefly review the key elements of the traditional Shannon-based MEP approach~\cite{Jaynes1957,Shore1980} to network ensembles, which leads to the standard ERG models~\cite{strauss1986general,anderson1999p,Holland1981,wasserman1994social,robins2007introduction,Park2004,bianconi2009entropy,fronczak2012exponential,Squartini2017,coolen2017generating,Cimini2019} that will be relevant as our benchmarks in subsequent sections. Our particular focus will be on how the Erd\H{o}s--R\'{e}nyi model~\cite{Erdoes1959,erdHos1960evolution} and the canonical Configuration Model~\cite{Park2003} naturally emerge in this context. More detailed expositions can be found, e.g. in Refs.~\cite{Park2004,Squartini2017}.

The MEP approach~\cite{Jaynes1957} focuses on Shannon's entropy $S[P]$~\cite{Shannon1948} which, in the context of networks, is defined in Eq.~\eqref{S1} over the set  $\mathcal{G}_{n}$ of all graphs with  $n$ nodes.
Here $P(G)$ represents the probability of graph $G$ in the ensemble. 
In the inductive inference framework, one encodes the available information in terms of $K$ canonical constraints $\{c_i(G)\}_{i=1}^K$, each of which is a function of the considered graph $G$.
For each constraint $c_i$, the available information is expressed in terms of a target value $c_i^*$ (generally, an observed value) and the constraining procedure consists of equating the expected value (ensemble average)  
\begin{equation}
    \langle c_i \rangle\equiv \sum_{G \in \mathcal{G}_{n}} P(G)\, c_i(G),\quad i=1,K\, ,
\label{eq:linearaverage}
\end{equation}
with the corresponding target value $c_i^*$. Note that we are using angular brackets to denote averages over $P(G)$. As an additional $(K+1)$-th constraint, one has to impose the normalization condition 
\begin{equation}
\sum_{G \in \mathcal{G}_{n}} P(G) = 1\, .
\label{eq:normalization}
\end{equation}

The resulting graph probability distribution $P(G)$ is obtained by maximizing the entropy functional, $S[P]$, subject to the aforementioned $K+1$ constraints. Introducing the associated $K+1$ Lagrange multipliers ${\bm{\theta}}=\{\theta_i\}_{i=1}^K$ and $\alpha$, the constrained maximization of $S[P]$ is achieved by imposing the vanishing of the partial derivatives of the Lagrange functional
\begin{eqnarray}
   \mathcal{L}[P] 
   \ &=& \ -\sum_{G \in \mathcal{G}_{n}}P(G) \ln P(G) \nonumber \\[2mm]  &&- \ \sum_{i=1}^{K}\theta_i \left(\sum_{G \in \mathcal{G}_{n}} c_i(G)P(G) - c^*_i\right)\nonumber \\[2mm]
   &&- \ \alpha\left(\sum_{G \in \mathcal{G}_{n}}P(G) - 1\right),\label{eq_lagrangian}
\end{eqnarray}
%
with respect to both the Lagrange multipliers and the values of $P(G)$.
The vanishing of the former derivatives generates the equations
\begin{equation}
    c^*_i\ \equiv \ \langle c_i \rangle,\quad i\ = \ 1,K\, ,
\label{eq:constraints}
\end{equation}
that enforce the desired constraints.
The vanishing of the latter derivatives leads to the functional form of $P(G)$ which, after eliminating the multiplier $\alpha$ via the normalization condition~\cite{Park2004,Squartini2017}, reads
\begin{eqnarray}
    P(G) \ = \  \frac{1}{\mathcal{Z({\bm{\theta}})}} \exp\left(-H(G,{\bm{\theta}})\right)\, ,
    \label{II.4.cf}
\end{eqnarray}
where%
\begin{equation}
    H(G,{\bm{\theta}}) \ =  \ \sum_{i=1}^{K} \theta_i c_i(G)\, ,
    \label{eq:graph_hamiltonian}
\end{equation}
is the so-called graph Hamiltonian.
 and
\begin{equation}
    \mathcal{Z}({\bm{\theta}}) \ = \ \sum_{G \in \mathcal{G}_{n}} e^{-H(G,{\bm{\theta}}) }\, ,
\end{equation}
is the ensuing partition function.
Therefore, in direct analogy with statistical physics, the form of $P(G)$ mimics the Boltzmann distribution for the canonical ensemble.

Once the probability $P(G)$ is determined, it can be used to express the expected value $\langle c_i \rangle$ of each constraint $c_i(G)$ as a function of the parameters $\bm{\theta}$.
Alternatively, the same result is obtained via the derivative
\begin{equation}
\langle c_i\rangle \ = \ -\frac{\partial}{\partial\theta_i}\ln\mathcal{Z}({\bm{\theta}})\, .
\end{equation}
Finally, the values of the parameters $\bm{\theta}$ are obtained by equating the expected value of each constraint with the corresponding target value into Eq.~\eqref{eq:constraints}, which in general represents a system of $K$ nonlinear coupled equations in the $K$ unknowns $\{\theta_i\}_{i=1}^K$.
Various techniques to identify the corresponding solution have been introduced~\cite{hunter2008ergm,snijders2002markov,coolen2017generating,vallarano2021fast}.

The central goal in the analysis of ERGs is the explicit calculation of the partition function $\mathcal{Z({\bm{\theta}})}$ for a given choice of the constraints $\{c_i(G)\}_{i=1}^K$. 
Depending on the definition of the latter, this task can be more or less easy (or even impossible)~\cite{Squartini2017,coolen2017generating}. 
For our purposes here, we will consider simple additive choices of the constraints (which are also the most widespread choices adopted in the analysis of real-world networks) that lead to explicit forms of $\mathcal{Z({\bm{\theta}})}$.
Importantly, these choices highlight the resulting independence between links in the graph, that follows from the Shannonian ($q=1$) assumption.
Therefore these models represent the cleanest benchmarks against which the novelty of $q$-ERGs will emerge in our subsequent analysis in Sec.~\ref{sec:qERGs}. 
Concretely, in the rest of the paper we consider examples of ensembles obtained by choosing constraints on simple undirected graphs (without self-loops) on $n$ nodes. From now on, $\mathcal{G}_{n}$ will therefore denote the set of all such graphs.
Note that the cardinality of this set is $|\mathcal{G}_{n}|=2^N$ where $N = \binom{n}{2} = n(n-1)/2$ is the number of distinct pairs of $n$ nodes (which coincides with maximum number of links that can be realized).

\subsection{Erd\H{o}s--R\'enyi model\label{sec:shannon_ER}}
%
We first consider the case where there is only one ($K=1$) constraint, namely the mean value of the total number of links in the graph, which we denote as $c_1(G)=L(G)$. 
It is convenient to introduce the adjacency matrix of a binary undirected graph $G$, which is a symmetric matrix $A$ with elements $a_{ij} = 1$ if there is a link between nodes $i$ and $j$, and $a_{ij} = 0$ otherwise (with $a_{ii} = 0$). For simple undirected graphs, the number of links in $G$ is given by 
\begin{equation}
L(G) \ = \  \sum_{i=1}^n\sum_{i<j} a_{ij}\, , 
\end{equation} 
and we set the expected constraint to the target value $c_1^*=L^*$ as in the general prescription given by Eq.~\eqref{eq:constraints}:
\begin{equation}
    L^* \ \equiv \ \langle L \rangle \ = \  \sum_{G \in \mathcal{G}_{n}} P(G) L(G)\, .
\label{eq_matchER}
\end{equation}
This leads to the graph Hamiltonian
\begin{equation}
    H(G,{{\theta}}) \ = \ \theta L(G)\ =\  \theta\sum_{i=1}^n\sum_{j\neq i} a_{ij}\, ,
    \label{eq:ER_hamiltonian}
\end{equation}
and partition function 
\begin{equation}\mathcal{Z}(\theta) \ = \ \sum_{G \in \mathcal{G}_{n}} e^{-\theta L(G)}.
\end{equation}
Changing notation from a sum over $G$ to a sum over all possible assignments $\{a_{ij}\}$ of the corresponding adjacency matrix entries, we can evaluate the partition function explicitly~\cite{Park2004,Squartini2017} as
\begin{eqnarray}
    \mathcal{Z}(\theta) &=& \sum_{\{a_{ij}\}} e^{-\theta \sum_{i=1}^n\sum_{j < i} a_{ij}} \ = \  \sum_{\{a_{ij}\}} \prod_{i=1}^n\prod_{j < i} e^{-\theta a_{ij}}\nonumber \\[2mm] &=& \prod_{i=1}^n\prod_{j < i} \sum_{a_{ij}=0,1} e^{-\theta a_{ij}} \ = \ \prod_{i=1}^n\prod_{j < i} (1 + e^{-\theta}) \nonumber \\[2mm] &=& (1 + e^{-\theta})^{N}\, .
\label{eq:erdos_renyi_partition_fn}
\end{eqnarray}
The ensuing probability of the entire graph $G$ then factorizes over pairs of nodes as follows:
\begin{eqnarray}
    P(G) &=& \frac{e^{-\theta \sum_{i=1}^n\sum_{j < i} a_{ij}}}{\mathcal{Z}(\theta)}  \nonumber \\[2mm] &=& \prod_{i=1}^n\prod_{j < i} \frac{e^{-\theta a_{ij}}}{1 + e^{-\theta}} \ = \  \prod_{i=1}^n\prod_{j < i} P(a_{ij})\, ,
    \label{10.aa}
\end{eqnarray}
where $P(a_{ij}) = {e^{-\theta a_{ij}}}/{(1 + e^{-\theta})}$ is the probability of the link between nodes $i$ and $j$ being either present ($a_{ij}=1$) or absent ($a_{ij}=0$)~\cite{Park2004,Squartini2017}. 

Thus, this model is a model of a random graph with independent links --- each link having the same probability
\begin{equation}
\langle a_{ij}\rangle \ = \  P(1) \ = \  \frac{e^{-\theta}}{1 + e^{-\theta}} \ = \  \frac{1}{1 + e^{\theta}}  \ \equiv \  p\, ,
    \label{eq:erdos_renyi_prob}
\end{equation}
of being present,  which is equivalent to modeling $\binom{n}{2}$ i.i.d. Boolean variables.
Such model of graphs with independent links, with each link having the same probability $p$, is precisely the celebrated {\em{Erd\H{o}s--R\'enyi}} (ER) model~\cite{Erdoes1959,erdHos1960evolution}. The connection probability $p$ is nothing but a reparametrization of the Lagrange multiplier $\theta$ that controls the expected total number of links 
\begin{equation}
\langle L\rangle \ = \ \sum_{i=1}^n\sum_{i<j}\langle a_{ij}\rangle \ = \ pN\, , 
\end{equation}
or, equivalently, the expected overall link density $\langle L\rangle/N=p$, see~\cite{Park2004,Squartini2017}.
Indeed, the value of $p$ enforcing the desired target number of links 
\begin{equation}
L^* \ = \ pN\, ,
\end{equation}
as required in Eq.~\eqref{eq_matchER}, is $p=L^*/N$, i.e. it equals the target link density.

The complete homogeneity of the ER model fails in replicating the complex features of real-world networks, such as the broad distribution of the number  of links (degree) per node and the asymptotic coexistence of \emph{sparsity} (vanishing link density) and \emph{clustering} (finite average fraction of realized triangles around nodes) for a large number $n$ of nodes.
For this reason, while the ER model remains a useful benchmark of homogeneous graphs, more complicated ERGs with node-specific constraints are used in network science, as we now consider in our next example.

\subsection{Canonical Configuration Model\label{sec:shannon_CM}}
%
In the canonical Configuration Model (CM), the expected degree (number of links) of each node is constrained~\cite{Park2004,Squartini2017}. The number of constraints is therefore $K=n$.
In terms of the adjacency matrix, the degree $k_i(G)$ of node $i$ in graph $G$ is defined as
\begin{equation}
k_i(G)\ =\  \sum_{j\neq i}a_{ij},\quad i=1,n\, .
\end{equation}
This leads to a graph Hamiltonian with $n$ constraints $c_i=k_i$ ($i=1,n$):
\begin{eqnarray}
    H(G,{\bm{\theta}}) &=& \sum_{i=1}^n\theta_i k_i(G)\ =\  \sum_{i=1}^n\sum_{j\neq i}\theta_i a_{ij} \nonumber \\[2mm]
    &=& \sum_{i=1}^n\sum_{j < i}(\theta_i + \theta_j) a_{ij}\, .
\label{eq:park_newman_hamiltonian}
\end{eqnarray}
The corresponding partition function is a straightforward generalization of Eq.~\eqref{eq:erdos_renyi_partition_fn} and is given by~\cite{Park2004,Squartini2017}
\begin{equation}
    \mathcal{Z}({\bm{\theta}}) \ = 
    \prod_{i=1}^n\prod_{j < i} \left(1 + e^{-\theta_i - \theta_j}\right)\, .
\label{eq:park_newman_partition_fn}
\end{equation}
The graph probability still factorizes as
\begin{equation}
    P(G)  \ = \ \prod_{i=1}^n\prod_{j < i} \frac{e^{-(\theta_i+\theta_j) a_{ij}}}{1 + e^{-(\theta_i+\theta_j)}} \ = \  \prod_{i=1}^n\prod_{j < i} P_{ij}(a_{ij})\, ,
    \label{PGCM}
\end{equation}
where $P_{ij}(a_{ij}) = {e^{-(\theta_i+\theta_j) a_{ij}}}/{(1 + e^{-(\theta_i+\theta_j)})}$ is the probability of the link between nodes $i$ and $j$ being either present ($a_{ij}=1$) or absent ($a_{ij}=0$). 
However, the link probabilities are now heterogeneous:
\begin{eqnarray}
    \langle a_{ij}\rangle  &=&   P_{ij}(1) \ = \   \frac{e^{-\theta_i-\theta_j}}{1 + e^{-\theta_i-\theta_j}}  \nonumber \\[2mm]  &=&  
    \frac{x_i x_j}{1+x_i x_j}  \ \equiv \ p_{ij}\, ,
    \label{eq:park_newman_prob}
\end{eqnarray}
where we have employed the conventional reparametrization from ${\bm{\theta}}$ to the so-called fitness parameters ${\bm{x}}$ via the prescription  $x_i = e^{-\theta_i}$~\cite{Park2004,Squartini2017}.

Now, equating the expected degrees to the target ones following the general requirement of Eq.~\eqref{eq:constraints} leads to the following $n$ coupled nonlinear equations for the $n$ unknown parameters~\cite{Squartini2011,Squartini2017}:
\begin{equation}
k_i^*\equiv \langle k_i\rangle \ = \  \sum_{j\neq i}p_{ij} \ = \   \sum_{j\neq i}\frac{x_i x_j}{1+x_i x_j},\quad i\ = \ 1,n\, ,
\label{eq_parametersCM}
\end{equation}
where $k_i^*$ is the desired target value of the degree of node $i$. Notably, these equations coincide with a maximum-likelihood condition on the parameters ${\bm{x}}$~\cite{Squartini2017}.
The value of ${\bm{x}}$ providing a solution to the above system of equations is unique, and efficient algorithms to find it for large networks where $k^*_i$ is prescribed for each $i$ separately (typically, when it represents the empirical degree observed in a real-world network~\cite{Squartini2011,Squartini2017}) exist~\cite{vallarano2021fast}. With these parameter values, the CM can be used as a null model of a given network with degrees $\{k^*_i\}_{i=1}^n$, e.g. to look for possible higher-order patterns in terms of statistically significant deviations from the maximum-entropy ensemble~\cite{Squartini2011,Squartini2017}.

There is however also a different, `reverse' approach, where the parameters ${\bm{x}}$ are treated as free parameters (e.g. the values $\{x_i\}_{i=1}^n$ can be sampled i.i.d. from a probability distribution of choice) that can be varied in order to explore the parameter space of the model and study the resulting value of the expected degrees~\cite{Park2004} and of other topological properties of the network, without a specific target network in mind.
Still, one might want to restrict the search in order to compare outcomes obtained by different values of ${\bm{x}}$ while keeping certain overall properties, such as the overall link density, fixed\footnote{Note that, in order to keep the expected overall link density fixed for different choices of the fitness vector ${\bm{x}}$, it is not appropriate to demand that the mean of the distribution from which the entries of ${\bm{x}}$ are sampled i.i.d. (hence the expected average of the $n$ fitness values) is kept fixed. This is due to the nonlinearity of the relation between fitness $x_i$ and expected degree $\langle k_i\rangle$ in Eq.~\eqref{eq_parametersCM}. For this reason, the approach described here and based on the global parameter $z$ is necessary.}. A particularly convenient strategy (that we will adopt later) is to rewrite $x_i\equiv\sqrt{z}s_i$ where $s_i$ is still a node-specific parameter, while $z$ is a global one. In this way it is possible to sample $\{s_i\}_{i=1}^n$ i.i.d. from any desired distribution, while adjusting $z$ to keep the average link density in the network fixed to a prescribed value, independent of that distribution. This means that the connection probability can be rewritten as 
\begin{equation}
    p_{ij} \ = \  \frac{z s_i s_j}{1  + z s_i s_j}\, ,
    \label{eq_FM}
\end{equation}
where $z$ is determined, conditionally on the realized values of $\{s_i\}_{i=1}^n$, by enforcing a constant expected link density $c^*=\langle L\rangle/N$ as follows:
\begin{equation}
c^* \ \equiv  \ \langle c\rangle \ = \   \frac{1}{N} \sum_{i=1}^n\sum_{j < i}\frac{zs_i s_j}{1+zs_i s_j}\, .
\label{eq_parametersFM}
\end{equation}
For fixed $\{s_i\}_{i=1}^n$, the above equation coincides with a maximum-likelihood condition for $z$~\cite{garlaschelli2008maximum}. With the above strategy, it is possible to vary $\{s_i\}_{i=1}^n$ and re-fit $z$ to effectively explore different fitness assignments $\{x_i\}_{i=1}^n$ that lead to exactly the same expected link density. This is also the standard approach in techniques of network reconstruction from partial infromation~\cite{Squartini2018,cimini2021reconstructing} that take $\{s_i\}_{i=1}^n$ as input in terms of some observable node feature, while requiring only the empirical knowledge of the overall link density $c^*$ in order to construct maximum-entropy ensembles for network inference based on link probabilities in the form of Eq.~\eqref{eq_FM}.
We will use the above approach to keep the network density fixed while exploring the $q$-exponential generalization of the CM in the next section.

The CM model improves upon the ER model by allowing the degree distribution of any real network to be replicated in expectation. However, it still cannot replicate the coexistence of sparsity (i.e. the average node degree remaining finite as the number $n$ of vertices becomes larger and larger) and clustering (i.e. a finite value of the local clustering coefficient). For this reason, models with independent links (like the ER and CM) are believed to be unable to replicate higher-order properties besides those included in the constraints.
This has led to the expectation that higher-order constraints need be included in the construction of the maximum-entropy probability in order to replicate those additional properties. However, in what follows we show that this conclusion is not correct when considering more general entropies. The same simple constraints as those defining the ER and CM can indeed lead to link dependencies, which in turn can generate higher-order properties, including a finite clustering in the sparse regime.

\section{$q$-exponential random graphs \label{sec:qERGs}}
%
We now come to our main contribution, by moving beyond the Shannon case considered in Sec.~\ref{sec:shannon} and considering the extended family of Uffink entropies given by Eq.~\eqref{uffink} as the object of the generalized MEP.
As already anticipated, the main result of the maximization of the Uffink functional will be the replacement of the inverse exponential distribution in Eq.~\eqref{II.4.cf} with the inverse $q$-exponential function defined in Eq.~\eqref{qexp}. 
This operation will define the new family of $q$-ERGs in terms of a certain distribution that we denote as $P_q(G)$.
To show explicitly how $P_q(G)$ emerges from the maximization of any of the entropies in Eq.~\eqref{uffink} for general $q$, we need to pick a specific choice of $f$.
Concretely, we select $f(x)=\ln_q(x)$ given by Eq.~\eqref{qlog}, which leads to the Tsallis entropy functional $T_q[P]$ in Eq.~\eqref{tsallis}~\cite{Tsallis1988}, even though the result will in the end not depend on $f$ (and on the resulting from of entropy) because, as we mentioned, the maximizer of a function is also the maximizer of any monotonic function $f$ of that function.

As in the Shannonian case, a key step is selecting the $K$ constraints $\{c_i(G)\}_{i=1}^K$ and imposing definite values for their expected values. However, while in the $q=1$ case Eq.~\eqref{eq:linearaverage} provides a unique natural definition of the expected value $\langle c_i\rangle$ of in terms of the distribution $P(G)$ itself, multiple definitions of $q$-expected values $\langle c_i\rangle_q$ exist in the $q\ne 1$ case, all reducing to the standard $\langle c_i\rangle$ in the limit $q\to 1$.  
In particular, three averaging procedures have been used more often in the context of $q$-exponentials. 
The so-called $q$-average, also known as the Curado--Tsallis weighted mean (or 2nd version of thermostatistics)~\cite{Curado1991}, replaces $P_q(G)$ with $P_q^q(G)$ in the definition of the expected constraints and offers certain formal advantages. 
For example, it allows one to set up  a formalism that closely emulates the formalism known from conventional thermodynamics~~\cite{Tsallis1998}. 
However, the $q$-average has a controversial status because it violates normalization ($\sum_G P_q^q(G)\ne 1$ for $q\ne 1$) and does not have an easy operational meaning. 
Another procedure is the 3rd version of thermostatistics, which replaces $P_q(G)$ with the so-called escort distribution $P_q^q(G)/\sum_{G'} P_q^q(G')$ and allows for a natural generalization of the maximum-likelihood principle to the estimation of the entropic parameter $q$~\cite{somazzi2025learn}, restoring the correspondence between entropy and negative log-likelihood. 
Conversely, the conventional linear averaging procedure based on $P_q(G)$ (also known as the Bashkirov's 1st version of thermostatistics) is conceptually cleanest from a statistical inference point of view~\cite{Jizba2019,Jizba2020}. 
In addition, it aligns well with network science, where standard (linear) ensemble averages are routinely used in ERGs as shown above. 
For this reason, we adopt standard averaging in our subsequent analysis and simply set 
\begin{equation}
\langle c_i\rangle_q \ \equiv \ \langle c_i\rangle \ = \ \sum_{G \in \mathcal{G}_{n}} P_q(G)\, c_i(G),\quad i \ = \ 1,K\, ,
\label{eq:qlinearaverage}
\end{equation}
in analogy with Eq.~\eqref{eq:linearaverage}. 
In any case, all the three choices (plus additional ones) can be made equivalent to each other by an appropriate change of $q$ (e.g. via the duality between $q$ and $2-q$) and/or the way the Lagrange multipliers $\{\theta_i\}_{i=1}^K$ are estimated~\cite{Ferri2005,somazzi2025learn}.

In order to maximize the entropy $T_q[P]$ under the $K$ constraints $\langle c_i\rangle=c^*_i$ ($i=1,K$) plus the additional constraint on the normalization of $P$ as in Eq.~\eqref{eq:normalization}, we define the $q$-Lagrange functional $ {\mathcal{L}}_q[P]$ which is a direct generalization of the Shannonian $ {\mathcal{L}}[P]$ in Eq.~\eqref{eq_lagrangian}:
\begin{eqnarray}
    {\mathcal{L}}_q[P] &=&  \frac{1}{q-1} \left(1 - \sum_{G \in \mathcal{G}_n} P^q(G)\right)\nonumber \\[2mm] &&- \ \sum_{i=1}^{K}\theta_i \left(\sum_{G \in \mathcal{G}_n} c_i(G)P(G) - c^*_i\right) \nonumber \\[2mm] &&- \ \alpha \left(\sum_{G \in \mathcal{G}_n} P(G) - 1\right).
    \label{eq:lagrange_functional}
\end{eqnarray}
Optimizing this functional (see Appendix~\ref{sec:qERG_derivation} for the full derivation) yields the $q$-exponential graph probability
%
\begin{eqnarray}
        P_q(G)      
         & =&  \frac{\left[ \exp_q\left(\frac{1}{q\mathcal{Z}_q^{1-q}({\bm{\theta}})}\sum_{i=1}^{K}\theta_i (c_i(G) - \langle c_i \rangle)\right) \right]^{-1}}{\mathcal{Z}_q({\bm{\theta}})} \nonumber\\
       &=& \frac{\exp_{2-q}\!\left(\frac{-1}{q\mathcal{Z}_q^{1-q}({\bm{\theta}})}\sum_{i=1}^{K}\theta_i (c_i(G) - \langle c_i \rangle)\right)}{\mathcal{Z}_q({\bm{\theta}})}\, ,~~~~~~ 
\label{eq:tsallis_model_general}
\end{eqnarray}
where the partition function equals $\mathcal{Z}_q(\bm{\theta})=U_q[P_q]$ and, in the second line, we have have exploited the convenient duality relation $\exp_q(-x)=\exp_{2-q}^{-1}(x)$ (see Appendix~\ref{sec:qERG_derivation}).

Although Eq.~\eqref{eq:tsallis_model_general} is self-referential (both the partition function $\mathcal{Z}_q$ and the constraints $\langle c_i\rangle$ appear inside the exponent), this problem can be resolved in a standard way by reparametrizing the Lagrange multipliers $\theta_i \to \tilde\theta_i$~\cite{Jizba2017} (see Appendix~\ref{sec:qERG_derivation} for details). Although the parameters $\tilde{\theta}_i$ are no longer the Lagrange multipliers of the original problem, they are still supposed to be fitted according to the constraints. To simplify notation, we henceforth drop the extra symbol and denote the redefined parameters simply as $\theta_i$ and the corresponding partition function as $\mathcal{Z}_q({\bm{\theta}})$. The general $q$-exponential probability can then be shown to take the compact form
\begin{equation}
    P_q(G) \ = \  \frac{1}{\mathcal{Z}_q({\bm{\theta}})} \exp_{2-q}\!\left(-H(G,{\bm{\theta}})\right),
    \label{eq:prob_with_hamiltonian}
\end{equation}
where $H(G,{\bm{\theta}})$ has the same definition as in Eq.~\eqref{eq:graph_hamiltonian} for the $q=1$ case. Equation~\eqref{eq:prob_with_hamiltonian} finally generalizes ERGs to a one-parameter family of $q$-ERGs, hence representing the formal solution to the most agnostic inferential construction of graph ensembles with soft constraints.

\subsection{Effective Hamiltonian}
\label{sec:effective_hamiltonian}
%
Clearly, when $q\to 1$ Eq.~\eqref{eq:prob_with_hamiltonian} reduces to the exponential distribution in Eq.~\eqref{II.4.cf}, and $q$-ERGs reduce to ordinary ERGs.
In order to better understand the differences (with respect to the Shannonian case) introduced by a value $q\ne 1$ in the $q$-exponential framework, we introduce the \textit{effective Hamiltonian} $H^{\text{eff}}_q(G,{\bm{\theta}})$ defined implicitly by
\begin{equation}
    P_q(G) \ \equiv \ \frac{1}{\mathcal{Z}_q({\bm{\theta}})}\exp\!\left(-H^{\text{eff}}_q(G,{\bm{\theta}})\right).
    \label{eq:prob_eff_hamiltonian}
\end{equation}
The quantity $H^{\text{eff}}_q(G,{\bm{\theta}})$ is defined in such a way to provide us with an indication of which `effective constraints' one should use in the Shannonian ERG formalism in order to produce a graph probability of the same form (up to a proportionality constant) as the ordinary exponential probability derived in Eq.~\eqref{II.4.cf} for the $q=1$ case. One might then see this operation as a mapping from the original set of constraints used to define a $q$-ERG to a different set of effective constraints necessary to define an equivalent ordinary ERG producing the same probability distribution $P_q(G)$.

Equation~\eqref{eq:prob_eff_hamiltonian} is equivalent to setting
\begin{eqnarray}
H^{\text{eff}}_q(G,{\bm{\theta}})&\equiv&-\ln \exp_{2-q}\!\left(-H(G,{\bm{\theta}})\right)\nonumber\\[2mm]
&=&\ln \exp_{q}H(G,{\bm{\theta}})\nonumber\\[2mm]
&=&\frac{1}{1-q}\ln \left[1 \ + \  (1 - q)H(G,{\bm{\theta}})\right]_+\nonumber\\[2mm]
&=&\sum_{k=1}^{+\infty}\frac{(q-1)^{k-1}}{k}H^k(G,{\bm{\theta}})\,.
    \label{eq:eff_hamiltonian}
\end{eqnarray}
where in the last line we require $|1 - q|\le |H(G,{\bm{\theta}})|^{-1}$ since we have employed the Mercator series expansion 
\begin{eqnarray}
    \ln(1+x) &=&\sum_{k=1}^{+\infty}\frac{(-1)^{k-1}}{k}x^k
\end{eqnarray}
of the logarithm, which is valid for $|x|\le 1$ and $x\ne -1$.
In particular, considering values of $q$ sufficiently close to $q=1$ (this is the most relevant regime, as we show in the next sections), we get the following expansion for the effective Hamiltonian:
\begin{eqnarray}
    H^{\text{eff}}_q(G,{\bm{\theta}}) &=& H(G,{\bm{\theta}}) \nonumber\\[2mm]
    && + \ \tfrac{1}{2}(1-q)H^2(G,{\bm{\theta}}) \nonumber\\[2mm]
    &&+\ \tfrac{1}{3}(1-q)^2 H^3(G,{\bm{\theta}}) \nonumber\\[2mm]
    &&+\ \tfrac{1}{4}(1-q)^3 H^4(G,{\bm{\theta}}) \nonumber\\[2mm]
    &&+\ \mathcal{O}\!\left((1-q)^4\right).
\end{eqnarray}
%

We thus realize that, as seen from an ordinary ERG viewpoint, the $q$-ERG generates the additional terms
\begin{eqnarray}
\Delta H_q(G,{\bm{\theta}})&\equiv&H^{\text{eff}}_q(G,{\bm{\theta}}) - H(G,{\bm{\theta}})\nonumber \\[2mm]
&=&\sum_{k=2}^{+\infty}\frac{(q-1)^{k-1}}{k}H^k(G,{\bm{\theta}})\, ,
\end{eqnarray}
besides those that would normally be present in the Shannon ERG case. These terms are produced by all possible $k$-th powers of $H(G,{\bm{\theta}})$ with $k>1$. It is however important to note that there are no additional Lagrange multipliers coupled with those extra terms: the only additional parameter is the single, global entropic parameter $q$, whose value couples with the $K$ native Lagrange multipliers ${\bm{\theta}}$ in determining $P_q(G)$ and automatically dictates the relative importance of any power $H^k(G,{\bm{\theta}})$ appearing in $H^{\text{eff}}_q(G,{\bm{\theta}})$.  Therefore the new effective quantities introduced by $\Delta H_q(G,{\bm{\theta}})$ should not in fact be interpreted as new tunable `constraints', but rather as new structural features that will be unavoidably promoted in the network ensemble, with a strength depending on the value of $q$, once the $K$ original constraints $\{\langle c_i\rangle\}_{i=1}^K$ are enforced.

A direct consequence of the presence of the extra ($k>1$) terms in the expansion~\eqref{eq:eff_hamiltonian} for $H^{\text{eff}}_q(G,{\bm{\theta}})$ is that, when the defining constraints in the Hamiltonian $H(G,{\bm{\theta}})$ of the $q$-ERG (i.e., the first term $k=1$) are sums of elements of the adjacency matrix (e.g. the total number of links or the degree sequence, as in the ER and CM examples we considered in Sec.~\ref{sec:shannon_ER} and~\ref{sec:shannon_CM} respectively), links will no longer be independent as we have found for the standard ERG, because the $q$-exponential of $H(G,{\bm{\theta}})$ generates effective higher-order link interactions at every order in $(q-1)$. Specifically, in such models the quadratic term $\frac{1}{2}(q-1)H^2(G,{\bm{\theta}})$ will introduce pairwise link--link interactions, the cubic term $\frac{1}{3}(q-1)^2 H^3(G,{\bm{\theta}})$ will introduce three-link interactions, and so on at every subsequent order ${k}^{-1}(q-1)^{k-1}H^k(G,{\bm{\theta}})$. Each such order involves products of $k$ link variables $a_{i_1 j_1} a_{i_2 j_2}\cdots a_{i_k j_k}$, thereby coupling links that are otherwise statistically independent in the ordinary ($q=1$) ERG case. Hence, \emph{even without modifying the constraints}, the mere departure from Shannon entropy ($q\neq 1$) transforms an edge-independent model into a genuinely higher-order one, coupling arbitrarily many links.
In the rest of the paper, we investigate in detail these $q$-ERGs and indeed find novel and much more general phenomenology with respect to what ordinary ERGs feature.

\section{$q$-exponential Erd\H{o}s--R\'enyi model \label{sec:qER}}
%
We now work out in detail the simplest possible $q$-ERG, which is the direct analogue of the ER model considered in Sec.~\ref{sec:shannon_ER}. 
The resulting model, which we denote as the $q$-exponential ER ($q$-ER) model, is still defined by the same single ($K=1$) constraint on the expected total number of links $\langle L\rangle$, which is set equal to a target value $L^*$, and the Hamiltonian is therefore the same as in Eq.~\eqref{eq:ER_hamiltonian}.
Using Eq.~\eqref{eq:prob_with_hamiltonian}, the resulting maximum $q$-entropy probability distribution is given by
\begin{eqnarray}
    P_q(G) \ &=& \ \frac{1}{\mathcal{Z}_q(\theta)} \exp_{2-q}\left(-\theta L(G)\right) \nonumber \\[2mm] &=& \frac{1}{\mathcal{Z}_q(\theta)} \exp_{2-q}\Big(-\theta \sum_{i=1}^n\sum_{j < i} a_{ij}\Big)\, ,
    \label{IV.30.kl}
\end{eqnarray}
with partition function
\begin{equation}
    \mathcal{Z}_q(\theta) = \ \sum_{\{a_{ij}\}} \exp_{2-q}\Big(-\theta \sum_{i=1}^n\sum_{j < i} a_{ij}\Big)\, .
    \label{IV.31.kl}
\end{equation}

The significant departure from the Shannonian case lies in the fact that, even though the Hamiltonian is still a sum of the pairwise elements $a_{ij}$, the maximum-entropy probability $P_q(G)$ no longer factorizes into a product of probabilities of individual links as in Eq.~(\ref{10.aa}), because the $\mathcal{Z}_q(\theta)$ does not factorize either: indeed, the $(2-q)$-exponential of the sum inside Eq.~\eqref{IV.31.kl} is not the product of the $(2-q)$-exponentials, unless $q=1$. 
This means that links are not independent random variables anymore, and the $q$-ER model is actually a higher-order network model with couplings among any number of links.
To confirm this, we notice that the effective Hamiltonian introduced in Eq.~\eqref{eq:prob_eff_hamiltonian} reads
\begin{eqnarray}
    H^{\text{eff}}_q(G,\theta) &=& H(G,\theta) \ + \ \Delta H_q(G,\theta)
    \label{eq:effective_hamiltonian_erm}\\[2mm]
    &=& \theta \sum_{i=1}^n\sum_{j < i} a_{ij}\nonumber\\[2mm]
    && +\ \tfrac{1}{2}(q-1)\theta^2 \left(\sum_{i=1}^n\sum_{j < i} a_{ij}\right)^{\!2} \nonumber\\[2mm]
    &&+\ \tfrac{1}{3}(q-1)^2 \theta^3 \left(\sum_{i=1}^n\sum_{j < i} a_{ij}\right)^{\!3} \nonumber\\[2mm]
    &&+\ \mathcal{O}\!\left((q-1)^3\right). \nonumber
\end{eqnarray}
We therefore see that each term of order $(q-1)^k$ generates a coupling between a number $k+1$ of pairs of nodes. 
In the next sections we will explore the consequences of these couplings and their relation with ERGs with higher-order constraints.

Before doing that, we notice that the (modified) Lagrange multipliers appearing in $P_q(G)$ are obtained by imposing the matching condition equivalent to the one in Eq.~\eqref{eq_matchER} with $P(G)$ replaced by $P_q(G)$, i.e.
\begin{equation}
    L^* \ \equiv \ \langle L \rangle \ = \ \sum_{G \in \mathcal{G}_{n}} P_q(G) L(G)\, ,
\end{equation}
where now the matching value of $\theta$ will depend on $q$~\cite{somazzi2025learn}, and will reduce to the ordinary value discussed in Sec.~\ref{sec:shannon_ER} when $q\to 1$.
Note that, even though links are all dependent on one another, the expected value $\langle a_{ij}\rangle$ still represents the probability of connection between nodes $i$ and $j$; however, this is now interpreted as the \emph{marginal} connection probability, unconditional on the realization of other links, which are all dependent on $a_{ij}$. Since $P_q(G)$ depends on the adjacency matrix $\{a_{ij}\}$ only through the total number of links $L(G) = \sum_{i<j} a_{ij}$, it is invariant under any permutation of links, so the value of $p_{ij}$ is independent of $i,j$, as in the traditional ER model:
\begin{equation}
  p_{ij} \ \equiv \ \langle a_{ij} \rangle \ = \ \frac{\langle L \rangle}{N} = \ \frac{L^*}{N}\, , \quad \;\;\;\forall i\ne j\, .
\end{equation}
The above relation is useful to compare the properties of the $q$-ER model with those of the corresponding ordinary ER model for the same expected link density ${L^*}/{N}$.

\subsection{Relation with ERGs with higher-order constraints}
\label{sec:relation}
We now go back to the effective Hamiltonian in Eq.~\eqref{eq:effective_hamiltonian_erm} and compare it with known ERGs with higher-order constraints. 

For instance, the term of order $q-1$ can be rewritten as $(\sum_{i=1}^n\sum_{j < i} a_{ij})^{2}=\sum_{i=1}^n\sum_{j < i} \sum_{k=1}^n\sum_{l < k} a_{ij}a_{kl}$, which generates 2-link couplings of the form $a_{ij}a_{kl}$. 
A subset of such couplings is present, for instance, in the so-called 2-star model~\cite{park2004solution,annibale2015two}, a popular ERG obtained by enforcing two global constraints in the network: the total number of links $\sum_{i=1}^n\sum_{j < i} a_{ij}$ (the lowest-order constraint) and the total number $\sum_{i=1}^n\sum_{j < i}\sum_{k < i} a_{ij}a_{ik}$ (with $k\ne j$) of so-called 2-stars, which is one of the simplest higher-order constraints. Each of the two constraints is controlled by a separate Lagrange multiplier.

Similarly, the term of order $(q-1)^2$ can be rewritten as $\sum_{i=1}^n\sum_{j < i} \sum_{k=1}^n\sum_{l < k} \sum_{u=1}^n\sum_{v < u}a_{ij}a_{kl}a_{uv}$, which generates 3-link couplings of the form $a_{ij}a_{kl}a_{uv}$. A subset of these couplings is present in the edge-triangle or Strauss model~\cite{strauss1986general,park2005solution,coolen2017generating}, another popular ERG obtained by enforcing the total number of links $\sum_{i=1}^n\sum_{j < i} a_{ij}$ and the total number $\sum_{i=1}^n\sum_{j < i}\sum_{k <j< i} a_{ij}a_{jk}a_{ki}$ of triangles. Here as well, the lower-order and the higher-order constraints are controlled by two independent Lagrange multipliers.

The 2-star model and the edge-triangle model are known to be characterized by a phase transition~\cite{strauss1986general,park2004solution,annibale2015two,park2005solution,coolen2017generating}: for certain combinations of the values of the two Lagrange multipliers, the ERG probability $P(G)$ is concentrated over a set of configurations where the constraints are close to their target expected values; however, for other combinations $P(G)$ is concentrated over multiple sets of configurations, each set featuring values of the constraints that are either much smaller or much larger than the target values, while still replicating the correct number in expectation. 
This result implies that, in individual realizations of the ERG, not all combinations of the values of the defining constraints can be realized. Certain combinations can only be replicated upon averaging over realizations that are individually very different from the desired configuration.

In the next sections we will show that a qualitatively similar phase transition is present in the $q$-ER model as well, however with important differences: unlike the 2-star model and the edge-triangle model, the $q$-ER model features couplings of all possible (infinite) orders; the introduction of these couplings is not \emph{ad hoc} or arbitrary, because the only chosen constraint is the total number of links, all the other quantities in $\Delta H_q(G,\theta)$ emerging spontaneously as a consequence of having relaxed the entropic parameter to values $q\ne 1$; the higher-order couplings need not be tuned independently of one another, since they are all governed by specific combinations of the two parameters $\theta$ and $q$.

\subsection{Thermodynamic limit}
\label{sec:qER_largeN}
%
We now study the asymptotic behavior of the $q$-ER probability distribution in Eq.~(\ref{IV.30.kl}) and the ensuing partition function for large $N$. The limit $N\to\infty$ corresponds to the `thermodynamic limit' where the size of the graph becomes infinite, while certain appropriately rescaled properties converge to limiting values.
%
%

We adopt a methodology familiar from the analysis of the Asymptotic Equipartition Property in information theory~\cite{Cover2006} and focus on the large-$N$ behavior of the quantity $(\ln P)/N$. To this end, we start with a related quantity known as the {\em free entropy density} 
\begin{equation}
    \Phi_q(\theta, N) \ = \ \frac{1}{N}\ln \mathcal{Z}_{q,N}(\theta)\, ,
    \label{eq:free_entropy_density}
\end{equation}
which an analogue of a Massieu function from equilibrium thermodynamics. 
%
From now on, we will use the notation $\mathcal{Z}_{q,N}(\theta)$, since we are interested in the large $N$ behavior and the partition function is explicitly $N$ dependent (in eq. \eqref{IV.31.kl} we are summing all the $N$ link variables $a_{ij}$, thus $N$ is effectively the system size). 
In appendix \ref{sec:appendix_qER_largeN}, we demonstrate that, in the limit of large $N$, the value of~(\ref{eq:free_entropy_density}) is determined by the maximum of yet another quantity, which we will refer to as the \textit{effective free entropy density} $\phi_q(c, \theta, N)$:
\begin{eqnarray}
    \lim_{N\to\infty} \Phi_q(\theta, N) \ = \lim_{N\to\infty} \max_{\ c \ \in \ [0,1]} \phi_q(c, \theta, N)\, ,
\label{eq:free_entropy_limit}
\end{eqnarray}
where
\begin{eqnarray}
&&\mbox{\hspace{-3mm}}\phi_q(c,\theta, N)\nonumber \\[2mm] &&\mbox{\hspace{5mm}}= \  H_b(c) \ - \  \frac{1}{(1-q)N}\ln\left[1 +  (1-q)N\theta c\right]\, ,~~~~~
    \label{eq:effective_free_entropy}
\end{eqnarray}
and $H_b(c) = -c \ln c - (1-c) \ln (1-c)$ is the binary entropy.

We now stress that, when considering the thermodynamic limit $N\to\infty$, one should crucially consider how the entropy parameter $q$ scales with $N$. We first consider $q$ constant and finite as $N$ grows, and show that this assumption has to be revised for any nontrivial result to emerge.

\subsubsection{Fixed $q$}
Assuming fixed $\theta$ and $q$ as $N\to\infty$, we have
\begin{equation}
    \lim_{N \to \infty} \frac{1}{(1-q)N} \ln\left( 1 + (1-q)N\theta  c \right) \ = \  0 \, .
\end{equation}
Thus, the second term in~(\ref{eq:effective_free_entropy}) becomes negligible for large $N$, and the behavior of $\Phi_q$ and $\mathcal{Z}_{q,N}(\theta)$ is determined entirely by the binary entropy term.
In the large $N$ limit (for fixed $\theta$ and $q$), we thus obtain for the free entropy density
\begin{equation}
    \lim_{N\to\infty} \Phi_q(\theta, N) \ = \ \max_{\ c \ \in \ [0,1]} H_b(c) \ = \  H_b(1/2)\, ,
    \label{45.kl}
\end{equation}
where we used the fact that the binary entropy $H_b(c)$ is maximized by $c_{\rm{max}} = \arg\max_c H_b(c) = \frac{1}{2}$.
Consequently, we have
\begin{eqnarray}
    ~~~~~&&\lim_{N\to\infty} \frac{\ln \mathcal{Z}_{q,N}(\theta)}{N} \ = \  H_b(1/2)\nonumber \\[2mm] &&~~~~~~~~\Leftrightarrow \quad \mathcal{Z}_{q,N}(\theta) \ \asymp \  e^{N H_b(1/2)} \ = \ 2^N\, ,~~~~~~~~
    \label{49.kl}
\end{eqnarray}
where we have employed the notation for an asymptotic equivalence in the exponential  sense, namely
\begin{eqnarray}
a_N \asymp b_N \; \ \Leftrightarrow \; \ \lim_{N\to\infty} \frac{1}{N}\ln a_N\ = \  \lim_{N\to\infty} \frac{1}{N}\ln b_N\, . ~~~~
\end{eqnarray}
%
%
Similarly, we also show in Appendix~\ref{sec:appendix_qER_largeN} that
\begin{eqnarray}
    P\!\left(\frac{L(G)}{N} = c\right) \ \asymp \  e^{-N\left[H_b(1/2) \ - \  H_b(c)\right]}\, .
    \label{eq:large_deviation_result}
\end{eqnarray}
%
%
This shows that the probability is maximal at \(c = 1/2\) and decays exponentially away from this value. Such behavior exemplifies the \textit{large deviation principle}~\cite{Touchette2009}, with the \textit{rate function} 
\begin{eqnarray}
I(c) \ \equiv \ H_b(1/2) \ - \  H_b(c)\, ,
\end{eqnarray}
determining the exponential decay. The unique minimum $I(1/2) = 0$ implies that only states with $c = 1/2$ retain non-negligible probability as $N \to \infty$, a phenomenon known as the \textit{concentration of measure}.

This reveals a striking \textit{degeneracy}: keeping $q$ and $\theta$ fixed (the latter condition corresponds to the dense graph case), the large-$N$ free entropy density converges to the universal value $H_b(1/2) = \ln 2$, identical to that of a completely unbiased Erd\"{o}s--R\'enyi graph with link probability $1/2$. In other words, both $\theta$ and $q$ become irrelevant in the thermodynamic limit at fixed values — the model loses all memory of its parameters and concentrates all probability on the maximally random configuration. This is clearly pathological from the modelling perspective, and motivates the question of whether and how this degeneracy can be overcome.

Let us now see how this degenerate case compares with the Shannonian one, where the probability of observing a graph with given density $c$ is given by
\begin{equation}
    P\left(\frac{L(G)}{N}  =   c\right) \ = \ \binom{N}{Nc}\ \! \frac{1}{\mathcal{Z}^N(\theta)} \ \! e^{-\theta Nc}\, .
\end{equation}
The effective free entropy density then takes the form 
\begin{eqnarray}
\phi_1(c, \theta, N) \ \equiv \  \phi_1(c, \theta) \ = \  H_b(c) \ - \  \theta c\, .
\label{eq_stamazza}
\end{eqnarray}
Notably, it is independent of $N$, if the contribution of the constraint term $\theta c$ remains finite (this is the dense graph limit serving as a reference for the dense $q\ne 1$ case considered above).  Extremizing with respect to $c$ yields a single stationary point at
\begin{eqnarray}
c_{\rm{max}} \ = \  \frac{1}{1 \ + \  e^\theta}\, ,
\end{eqnarray}
in agreement with eq. \eqref{eq:erdos_renyi_prob}. The probability then reads
\begin{equation}
    P\left(\frac{L(G)}{N} = c\right) \ \asymp \  e^{-N\left[\phi_1(c_{max}, \theta) \ - \  \phi_1(c, \theta)\right]}\, .
\end{equation}
Note that if we choose $\theta=0$, we get $\phi_1(c, \theta) = H(c)$ and $c_{\rm{max}} = 1/2$, which corresponds to our results for the $q$-exponential model. This means that in the large $N$ limit, the $q$-exponential model behaves the same as the ($1$-exponential) ER model with $\theta=0$ or equivalently with a link probability $p=1/2$.

\subsubsection{Scaled $q$\label{IV.C.kk}}

As shown in our previous analysis, the entire large-$N$ behavior of the $q$-exponential model is encoded in the function $\phi_q(c, \theta, N)$ appearing in Eq.~\eqref{eq:effective_free_entropy}, as a consequence of the concentration of measure. We have just shown that, in the large $N$ limit, $\phi_q(c, \theta, N)$ approaches a trivial limit regardless of $q$, if the latter remains fixed as $N$ grows. We now show that this degeneracy disappears if, assuming the `dense' regime where $\theta$ remains finite in the thermodynamic limit, we let $q$ scale with $N$ as 
\begin{equation}
q \ = \ 1 \ - \ \frac{r}{N}\, ,
\label{eq_scaledq}
\end{equation}
where $r$ is a finite constant (independent of $N$), which implies the finite limit
\begin{equation}
\lim_{N\to \infty}(1-q) N \ = \  r\, .
\end{equation} 
With this scaling, the second term in~Eq.\eqref{eq:effective_free_entropy} remains finite and provides a new important contribution, genuinely inherent to the $q$-exponential model. 
We also see from Eq.~\eqref{eq:effective_hamiltonian_erm} that the chosen scaling for $q$ implies that all higher-order terms in the effective Hamiltonian become of the same order $\mathcal{O}(N)$ as $H(G,\theta)$, so that the $q$-ER model will take into account all possible $k$-link couplings (with $k=1,\dots,\infty$), each of them contributing with the same relevance. 
%

To pave the way for the investigation of the effects of the scaling in Eq.~\eqref{eq_scaledq}, we note again that, for large $N$, the probability distribution becomes sharply peaked around the configurations that maximize the effective free entropy density $\phi_q(c, \theta, N)$. In this regime, we can carry out a \textit{saddle-point approximation} and replace the ensemble average with the most likely value. 
With the reparametrization in Eq.~\eqref{eq_scaledq}, the effective free entropy density takes the form
\begin{equation}
    \phi_r(c, \theta) \ = \  H_b(c) \ - \ \frac{1}{r}\ln(1+r\theta c)\, ,
\end{equation}
which reduces to its Shannonian counterpart in Eq.~\eqref{eq_stamazza} as $r\to0$.
To find the maxima, we compute the derivative of $\phi_r(c, \theta)$ with respect to $c$, which reads
\begin{equation}
    \frac{\partial}{\partial c} \phi_r(c, \theta) \ = \ \ln\left(\frac{1-c}{c}\right) \ - \  \frac{\theta}{1+r\theta c}\, .
    \label{eq:saddle_point_equation_derivative}
\end{equation}
Setting this derivative to zero leads to the \textit{saddle-point equation}
\begin{equation}
    c \ =  \ \frac{1}{1 \ + \  \exp\left(\frac{\theta}{1+r\theta c}\right)}\, .
    \label{eq:saddle_point_equation}
\end{equation}
We have seen in Eq.~\eqref{eq:erdos_renyi_prob} that in the Shannonian case $(r=0)$, there is a single solution at~$c = 1/(1+e^\theta)$. For~$r\neq 0$, however, the behavior can be different, as we show in the next section.

\subsection{Phase transition}
\label{sec:phase_transition}
The saddle-point equation~\eqref{eq:saddle_point_equation} allows us to characterize the effects of the link dependencies implied by the chosen scaling for $q$, showing that they lead to a phase transition similar to the one observed for traditional ERGs with higher-order constraints, even if the $q$-ER still enforces only a single constraint on the total number of nodes.

\begin{figure}[b]
    \centering
    \includegraphics[width=0.47\textwidth]{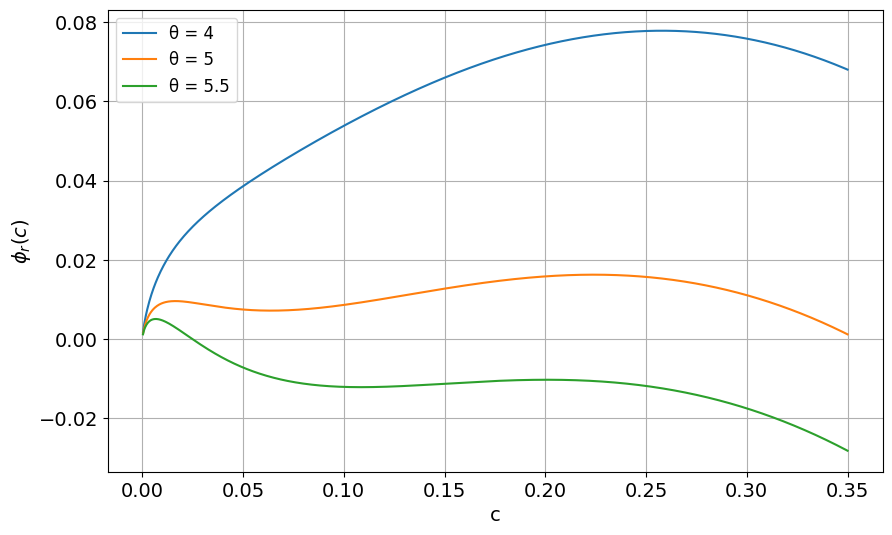}
    \caption{The effective free entropy density $\phi_r(c,\theta)$ for $r=2.7$ as a function of link density $c$ for different values of $\theta$. The global maxima of these curves determine the typical configuration.}
    \label{fig:free_entropy_theta}
\end{figure}

Indeed, Fig.~\ref{fig:free_entropy_theta} shows the effective free entropy density~$\phi_r(c, \theta)$ for~$r = 2.7$ at~$\theta = 4, 5, 5.5$. For~$\theta = 4$, a unique maximum is observed, whereas for~$\theta = 5$ and~$\theta = 5.5$ two local maxima emerge. Furthermore, the global maximum shifts abruptly between~$\theta = 5$ and~$\theta = 5.5$, suggesting the occurrence of a phase transition.

\begin{figure*}[tbp]
    \centering
    \includegraphics[width=0.96\textwidth]{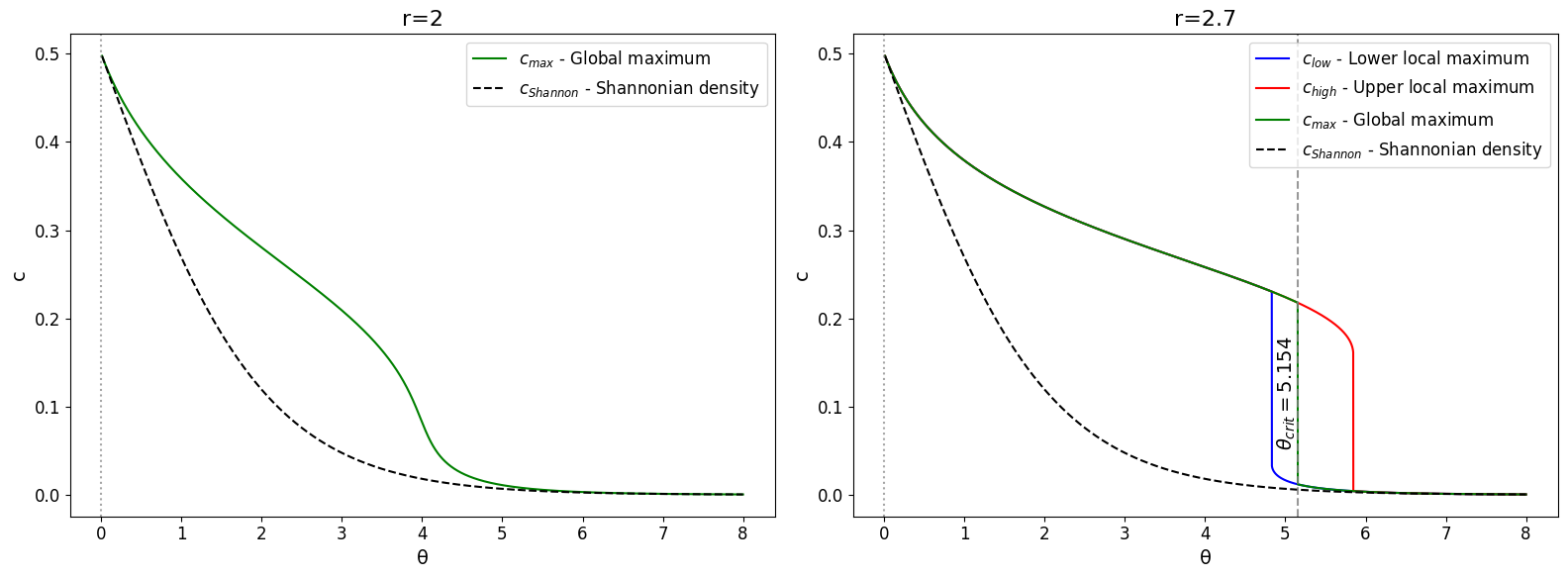}
    \caption{On the left, we show the behavior of the typical configuration link density $c_{max}$ for $r=2$ as a function of $\theta$. On the right, for $r=2.7$, we plot the local and global maxima of the effective free entropy density $\phi_r(c,\theta)$ as a function of $\theta$. In the range $\theta\in[4.831,5.846]$, there are two local maxima $c_{low}$ and $c_{high}$, where one is stable and one metastable. The phase transition happens at $\theta_{crit} \approx 5.154$. For reference, we also show the Shannonian network density $c_{Shannon} = {(1+e^\theta)^{-1}}$.}
    \label{fig:phase_transition_network}
\end{figure*}

To confirm whether a phase transition occurs, we compute the local maxima and identify the global maximum as a function of~$\theta$. In Fig.~\ref{fig:phase_transition_network}~(right), we show the locations of the local maxima as a function of~$\theta$. For~$r = 2.7$, two distinct local maxima, denoted~$c_{\text{low}}$ and~$c_{\text{high}}$, exist within the interval~$\theta \in [4.831, 5.846]$, corresponding to a stable and a metastable state, respectively. The points~$\theta_{sp-} \approx 4.831$ and~$\theta_{sp+} \approx 5.846$ are the \textit{spinodal points}. The phase transition occurs at~$\theta_{\text{crit}} \approx 5.154$, below which the high-density state is stable and above which the low-density state becomes stable. For comparison, Fig.~\ref{fig:phase_transition_network}~(left) shows the case~$r = 2$, where no phase transition is observed. The Shannonian network density, given by~$c_{\text{Shannon}} = 1 / (1 + e^{\theta})$, is also shown for reference.

The conditions for the phase transition can be made precise. In Appendix~\ref{sec:phase_transition_condition}, we establish the following result:
\newtheorem{proposition}{Proposition}
\begin{proposition}
Let $x^*>0$ denote a solution of the equation
\begin{equation}
    \tanh\!\left(\frac{1}{x^*}\right) \ = \  x^* \,.
\end{equation}
The phase transition in the $q$-ER model occurs if and only if
\begin{equation}
    r \ \equiv \  (1-q)N \ > \ r_{\mathrm{cp}}\ \equiv \ \frac{(x^*)^2}{1-(x^*)^2} \ \approx \ 2.2768 \,.
\end{equation}
\end{proposition}
The critical value $r_{\mathrm{cp}} \approx 2.2768$ is thus the threshold below which no phase transition can occur. At $r = r_{\mathrm{cp}}$, the free entropy density $\phi_{r_{\mathrm{cp}}}(c,\theta)$ first develops an inflection point as $\theta$ is varied; we define $\theta_{\mathrm{cp}}$ as this threshold value. 

In Appendix~\ref{sec:critical_point}, we establish the following proposition:
\begin{proposition}   The critical point $\theta_{\rm{cp}}$ is given by
    \begin{equation}
        \theta_{\rm{cp}} \ = \ 2 \ + \  \sqrt{\frac{4(r_{cp}+1)}{r_{\rm{cp}}}} \ \approx \  4.3994\, .
    \end{equation}
\end{proposition}

\begin{figure}[b]
    \centering
    \includegraphics[width=0.47\textwidth]{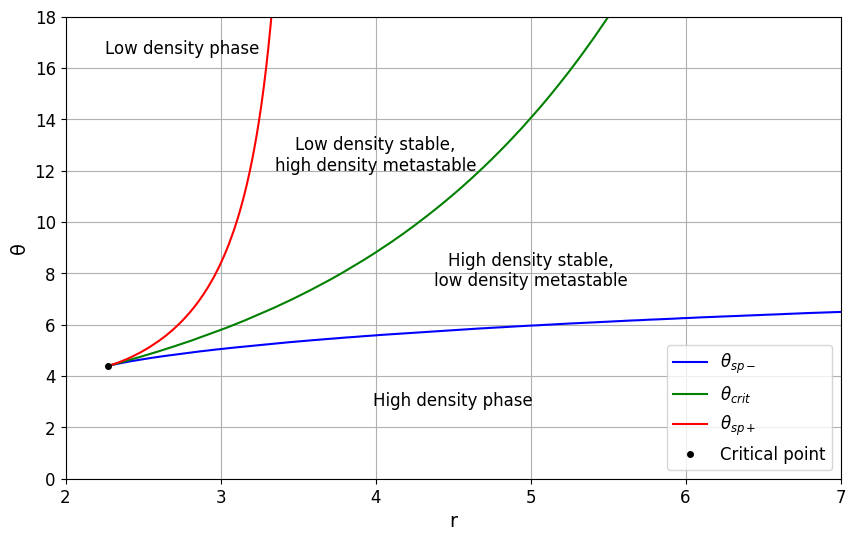}
    \caption{The phase diagram of the $q$-ER model. For each value of $r$, we show the spinodal points $\theta_{\rm{sp}-}$ and $\theta_{\rm{sp}+}$ as well as the phase transition $\theta_{\rm{crit}}$. The critical point $(r_{\rm{cp}}, \theta_{\rm{cp}}) \approx (2.2767, 4.3994)$ is shown as the black dot.}
    \label{fig:phase_diagram}
\end{figure}

We map out the full phase diagram by numerically solving the saddle-point equation~\eqref{eq:saddle_point_equation} for a range of values of $r$. For each value we determine the two spinodal points $\theta_{\rm{sp}-}$ and $\theta_{{\rm sp}+}$, the phase-transition boundary $\theta_{\text{crit}}$, and the critical point $(r_{\rm{cp}},\,\theta_{\rm{cp}}) \approx (2.2767,\,4.3994)$. The result is presented in Fig.~\ref{fig:phase_diagram}.


\begin{figure}[b]
    \centering
    \includegraphics[width=0.47\textwidth]{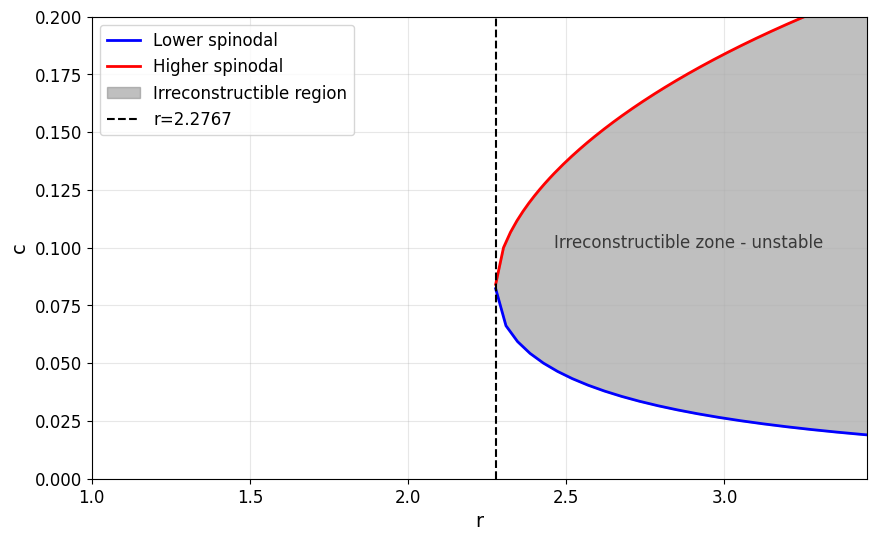}
    \caption{For different values of $r$, we numerically approximated boundaries of the region, where density reconstruction in the $q$-ER model is not possible. Such a region is determined by the lowest possible value of $c_{\rm{high}}$ and the highest possible value of $c_{\rm{low}}$ in the metastable regime. We also show $r$ value of the critical point, $r_{\rm{cp}}=2.2767$.}
    \label{fig:reconstruction_boundaries}
\end{figure}

Finally, we note that the presence of a phase transition has direct implications for practical applications, such as network reconstruction. 
In particular, for a given value of $r$ there may be no choice of $\theta$ that reproduces a prescribed target network density. 
In the metastable region, two distinct phases coexist --- a low-density phase $c_{\text{low}}$ and a high-density phase $c_{\text{high}}$. 
Consequently, densities lying between maximum value of $c_{\text{low}}$ and minimal value of $c_{\text{high}}$ cannot be reconstructed. 
This behavior is already evident in Fig.~\ref{fig:phase_transition_network}. 
For a range of $r$-values, we numerically determined these boundaries and identified the region in which density reconstruction is impossible. 
The results are shown in Fig.~\ref{fig:reconstruction_boundaries}. 
We further assume that the irreconstructible region increases with $r$, and in the large-$r$ limit, only allows either $c=0$ or $c=0.5$.

Remarkably, we have therefore found behavior typical of standard ERGs defined via the addition of higher-order constraints besides the overall number of links~\cite{strauss1986general,park2004solution,annibale2015two,park2005solution,coolen2017generating,fronczak2007phase}. However, in our $q$-ERG setting that behaviour does not actually require such extra constraints in the definition of the Hamiltonian. It is the regime $q\ne 1$ that automatically produces the phenomenology typical of higher-order ERGs, while keeping the overall link density as the only target constraint.

\subsubsection{Note on the order of limits }
%

In statistical physics, phase transitions are typically characterized by the non-analytic behavior of thermodynamic potentials, such as the Helmholtz or Gibbs free energy. 
Such non-analyticities, however, arise only in the thermodynamic limit $N \to \infty$. 
Consider again Eq.~\eqref{eq:free_entropy_limit}: indeed, the non-analyticity is caused by the maximization over $c$ in the limit $N\to\infty$.
However, we previously argued that in this limit the model approaches the Shannonian regime with $\theta = 0$, for which no phase transition is present. 

This apparent contradiction arises from the fact that we imposed $r = (1-q)N$ and treated $r$ as fixed. Doing so implicitly changes the structure of the limiting procedures. If we first take the limit $N \to \infty$ at fixed $q$, we indeed recover the degenerate Shannonian model in which the influence of the parameter $\theta$ vanishes, as previously discussed. Taking the subsequent limit $q \to 1$ does not alter this conclusion.
Conversely, if we instead take the limit $q \to 1$ first, we recover the standard Erdős--R\'enyi model. This \emph{non-commutativity of limits} was already emphasized in~\cite{Hanel2011}, where it was shown that Tsallis entropy lie in the $(c,d) = (q,0)$ universality class, whereas Shannon and R\'enyi entropies correspond to $(c,d) = (1,1)$. These belong to different scaling classes and therefore exhibit distinct behaviors in the thermodynamic limit $N \to \infty$.  The situation is also reminiscent of the standard theory of phase transitions, where 
\begin{eqnarray}
\lim_{N \rightarrow \infty} \lim_{T \rightarrow T_c} \ \neq \  \lim_{T \rightarrow T_c} \lim_{N \rightarrow \infty}\, ,
\end{eqnarray}
with $T_c$ denoting the critical temperature of the phase transition.

Our central observation is that the $N\to\infty$ limit taken with $(1-q)N$ held constant yields nontrivial behavior, and that a phase transition emerges only under this scaling. Moreover, we observe signatures of this transition to persist even at finite $N$, manifesting, for example, through metastable states. In network ensembles, where $N$ (the number of possible links) is typically large but fixed, the saddle-point approximation should therefore remain applicable.

Let us now examine the implications of the scaling $r = (1 - q)N$ for the entropy. In particular, the Tsallis entropy may be equivalently rewritten in terms of the parameter $r$ as  
\begin{equation}
    T_r[P] =  \frac{N}{r}\left(1-\sum_{G\in\mathcal{G}_n} [P(G)]^{r/N + 1}\right).
\end{equation}
By considering the uniform distribution $P_u(G)=1/W$ for all $G\in\mathcal{G}_n$, where $W=2^N$ is the number of possible binary undirected graphs with $n$ nodes and $N=n(n-1)/2$ node pairs, we obtain
\begin{equation}
    T_r[P_u] \ = \   \frac{N}{r}\left(1-{W^{-r/N}}\right)\ = \ N\frac{1-2^{-r}}{r}\propto \  N\, ,
\end{equation}
meaning that the entropy is extensive! So, by the specific choice of the limiting scaling $\lim_{N\to\infty}(1-q)N = r$ with $r$ finite, we actually made the Tsallis entropy extensive again. However, remarkably, the resulting behavior is much richer than in the Shannonian case, which (for a given exponential state-space scaling) is extensive as well.

\subsection{Moments and cumulants}
\label{sec:Erdos_Renyi_moments_and_cumulants}
%
The saddle-point approximation identifies the typical link density in the large-$N$ limit, but does not capture fluctuations or correlations between links. To access these, we now compute the exact moments and cumulants under the $q$-exponential model, starting with the average number of links. In Shannonian case we know that
\begin{equation}
    \langle L \rangle \ = \ - \frac{\partial \ln \mathcal{Z(\theta)}}{\partial \theta}\, .
    \label{IV.58.kl}
\end{equation}
This equation, however, does not hold in the $q$-exponential case. In particular,  in Appendix~\ref{sec:average_links}, we show that instead we have
\begin{equation}
    \langle L \rangle \ = \  N\exp_{2-q}(-\theta)\frac{\mathcal{Z}_{q,N-1}\!\left(\frac{\theta}{1-(q-1)\theta}\right)}{\mathcal{Z}_{q,N}(\theta)}\, .
    \label{eq:avg_num_links}
\end{equation}

Note, in particular, the appearance of
\begin{eqnarray}
\mathcal{Z}_{q,N-1}\!\left(\frac{\theta}{1-(q-1)\theta}\right)\, ,
\end{eqnarray}
i.e., the partition function for a system with one fewer link in the sum, and with a modified parameter.  Thus, the general relationship is not expressed via a derivative, but through an explicit formula connecting partition functions with a different number of summed links.  It might be easily checked that, as $q\to1$, Eq.~(\ref{eq:avg_num_links}) reduces to Eq.~(\ref{IV.58.kl}).

The marginal probability of a single link is, due to permutation symmetry of the probability, just $\langle a_{ij} \rangle = \langle L \rangle / N = \langle c \rangle$ for all $(i,j)$, i.e. it is the same as the average link density. We therefore have
\begin{equation}
    \langle a_{ij} \rangle \ = \ \langle c \rangle \ = \ \exp_{2-q}(-\theta)\frac{\mathcal{Z}_{q,N-1}\!\left(\frac{\theta}{1-(q-1)\theta}\right)}{\mathcal{Z}_{q,N}(\theta)}\, .
    \label{eq:avg_link_density}
\end{equation}
In Appendix~\ref{sec:direct_computation_of_avg_link_density}, we show that, instead of computing $\langle a_{ij} \rangle$ using $\langle L \rangle$, there is actually an elegant way to compute it directly. This is because one can easily sum over other degrees of freedom. The same methodology can be extended to the computation of higher-order joint moments. For $k$ distinct links, the following identity is derived in Appendix~\ref{sec:higher_moments}:
\begin{eqnarray}
    \langle a_{i_1 j_1} \cdots a_{i_k j_k} \rangle =   \exp_{2-q}(-k\theta) \frac{\mathcal{Z}_{q,N-k}\left(\frac{\theta}{1-k(q-1)\theta}\right)}{\mathcal{Z}_{q,N}(\theta)}\, .~~~
    \label{eq:tsallis_higher_moments}
\end{eqnarray}

Although the above expression is elegant, it presents a computational challenge due to the difficulty of evaluating the partition function. Its characteristic asymptotic behavior is known, see Eq.~(\ref{49.kl}), however, its exact value remains unknown, even for large $N$. Since, as discussed above, a particularly relevant regime corresponds to values of $q$ close to one, we now turn to a perturbative analysis of this case.

In terms of the effective Hamiltonian $H^{\text{eff}}_q(G,\theta)$ introduced in Eq.~\eqref{eq:effective_hamiltonian_erm}, the partition function is given by
\begin{eqnarray}
    \mathcal{Z}_{q,N} \ = \  \sum_{G\in\mathcal{G}_N} \exp\!\left(-H^{\text{eff}}_q(G,\theta)\right)\, .
\end{eqnarray}
By introducing source terms $h_{ij}$ 
coupled to individual links, we can then derive the standard identities
\begin{eqnarray}
   &&\mbox{\hspace{-2mm}} \mathcal{Z}_{q,N}(\theta, \{h_{ij}\})\nonumber \\[2mm] 
   &&\mbox{\hspace{7mm}}= \sum_{G\in\mathcal{G}_N} \exp\!\Big(-H^{\text{eff}}_q(G) - \sum_{i=1}^n\sum_{j < i} h_{ij} a_{ij}\Big) \, ,~~~~\label{83.kl}
   \end{eqnarray}
   and
   \begin{eqnarray}
   \frac{\partial \ln \mathcal{Z}_{q,N}}{\partial h_{kl}}\bigg|_{\{h_{ij}\}=0} &=& \langle a_{kl} \rangle\, , \label{eq:cumulant_first_derivative} \\[2mm]
    \frac{\partial^2 \ln \mathcal{Z}_{q,N}}{\partial h_{kl} \partial h_{uv}}\bigg|_{\{h_{ij}\}=0} &= &\langle a_{kl} a_{uv} \rangle \ -\  \langle a_{kl} \rangle \langle a_{uv} \rangle \, ,
    \label{eq:cumulant_second_derivative}
\end{eqnarray}
the last expression coinciding with the covariance $\mbox{Cov}(a_{kl}, a_{uv})$.
By isolating the Shannon-entropy contribution from the effective Hamiltonian, we can write
\begin{eqnarray}
    \mathcal{Z}_{q,N}(\theta, \{h_{ij}\}) &=& \sum_{\{a_{ij}\}} \exp\!\Big[-\sum_{i=1}^n\sum_{j < i} (\theta - h_{ij}) a_{ij}\Big] \nonumber\\[2mm]
    &&\times\ \exp\!\left(-\Delta H_q(G)\right)\nonumber \\[3mm]
    &=& \mathcal{Z}_{1,N}(\theta, \{h_{ij}\}) \  \langle \exp(-\Delta H_q) \rangle_{\text{Sh}}\, ,~~~~~
\end{eqnarray}
where $\langle \cdots \rangle_{\text{Sh}}$ denotes the ensemble average with respect to the Shannonian distribution, and $\Delta H_q(G)$ denotes the non-Shannonian part of the effective Hamiltonian, which can be expanded as
\begin{eqnarray}
    \Delta H_q(G) &=& \tfrac{1}{2}(q-1)\theta^2 L(G)^2 \ +\ \tfrac{1}{3}(q-1)^2 \theta^3 L(G)^3 \nonumber\\[3mm]
    &+& \tfrac{1}{4}(q-1)^3 \theta^4 L(G)^4 \ +\ \mathcal{O}\!\left((q-1)^4\right).
\end{eqnarray}\\
To compute the cumulants, we need the logarithm of the partition function, i.e.
\begin{eqnarray}
    \ln \mathcal{Z}_{q,N}(\theta, \{h_{ij}\}) &=& \ln \mathcal{Z}_{1,N}(\theta, \{h_{ij}\}) \nonumber\\[2mm]
    &&+ \  \ln \langle \exp(-\Delta H_q) \rangle_{\text{Sh}}\nonumber \\[2mm]
    &=& \prod_{i=1}^n\prod_{j < i} \ln(1+e^{-\theta + h_{ij}}) \nonumber\\[2mm]
    &&+ \  \ln \langle \exp(-\Delta H_q) \rangle_{\text{Sh}}\, .
    \label{eq:ln_partition_fn}
\end{eqnarray}

We  can then use the cumulant expansion
\begin{widetext}
\begin{eqnarray}
\ln \langle \exp(-\Delta H_q) \rangle_{\text{Sh}} &=& - \langle \Delta H_q \rangle_{\text{Sh}} +  \frac{1}{2} \Big(\langle \Delta H_q^2 \rangle_{\text{Sh}} \ - \  \langle \Delta H_q \rangle_{\text{Sh}}^2\Big) +  \mathcal{O}\!\left((q-1)^3\right) \nonumber\\[2mm]
&=& -\frac{1}{2}(q-1)\theta^2 \langle L^2 \rangle_{\text{Sh}} +  (q-1)^2\left[-\frac{1}{3}\theta^3 \langle L^3 \rangle_{\text{Sh}} + \frac{1}{8}\theta^4 \left(\langle L^4 \rangle_{\text{Sh}} - \langle L^2 \rangle_{\text{Sh}}^2\right)\right] +  \mathcal{O}\!\left((q-1)^3\right)\, .~~~~
\end{eqnarray}
%
To extract the cumulants, we differentiate $\ln \mathcal{Z}_{q,N}$ with respect to the source terms $h_{ij}$ as prescribed by~\eqref{eq:cumulant_first_derivative} and~\eqref{eq:cumulant_second_derivative}. These computations are carried out in Appendix~\ref{sec:q_approx_1_num_links}, yielding, to first order in $(q-1)$:
\begin{eqnarray}
    &&\langle a_{ij} \rangle  \ = \  \frac{1}{1+e^\theta}\left[1 \ + \ \frac{1}{2}(1-q)\theta^2\frac{e^\theta}{(1+e^\theta)^2}(2N - 1 + e^\theta) \right]\ + \  \mathcal{O}((q-1)^2)\, , \label{eq:er_marginal_link_probability}\\[2mm]
    &&\mbox{Cov}(a_{ij}, a_{kl}) \ = \ (1-q)\frac{\theta^2e^{2\theta}}{(1+e^\theta)^4}  \ + \  \mathcal{O}((q-1)^2)\, .
    \label{eq:er_edge_covariance}
\end{eqnarray}

As expected from the higher-order link interactions induced by $q \neq 1$, the covariance is positive for $q < 1$ and vanishes at leading order as $q \to 1$. Notably, to first order in $(q-1)$ it is independent of $N$, though this independence breaks down at higher orders.

\begin{figure*}[t]
    \centering
    \includegraphics[width=0.98\textwidth]{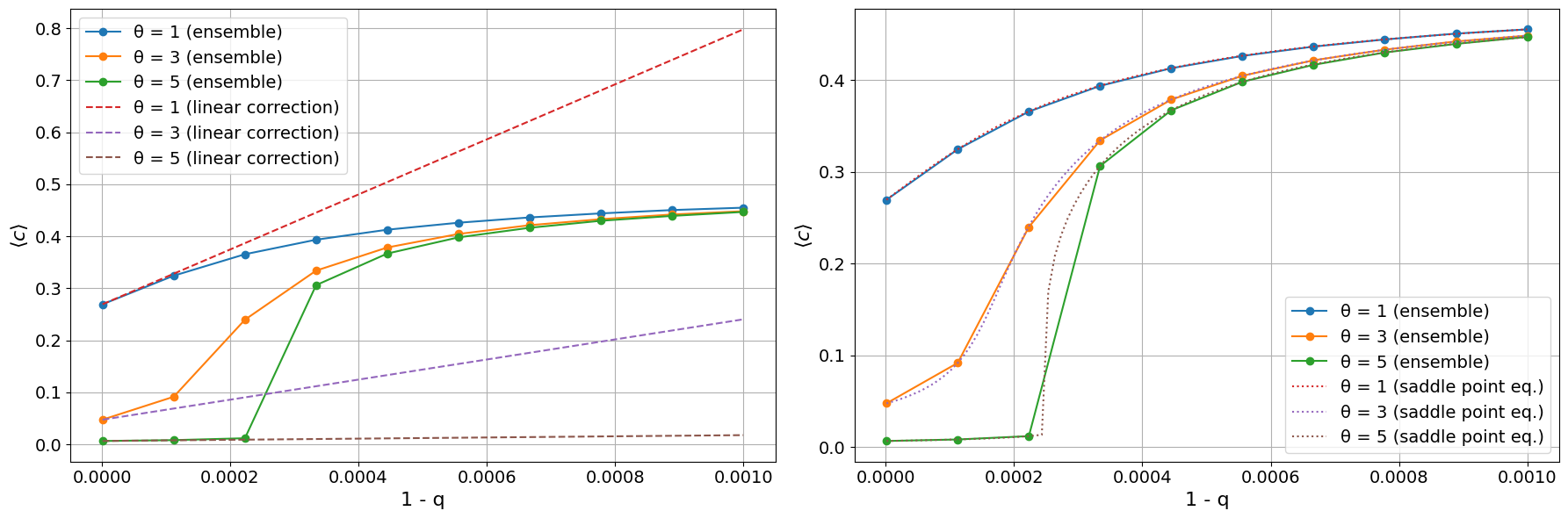}
    \caption{For ensembles of 1000 graphs with 100 nodes, we compute the average link density $\langle c \rangle$ and compare with the perturbative linear correction [Eq.~\eqref{eq:er_marginal_link_probability}] and with the solutions of the saddle-point equation [Eq.~\eqref{eq:saddle_point_equation}, on the right].}
    \label{fig:link_densities}
\end{figure*}

\end{widetext}

\subsection{Network properties}\label{sec:qERM_network_properties} 

\begin{figure*}[!t]
    \centering
    \includegraphics[width=0.63\textwidth]{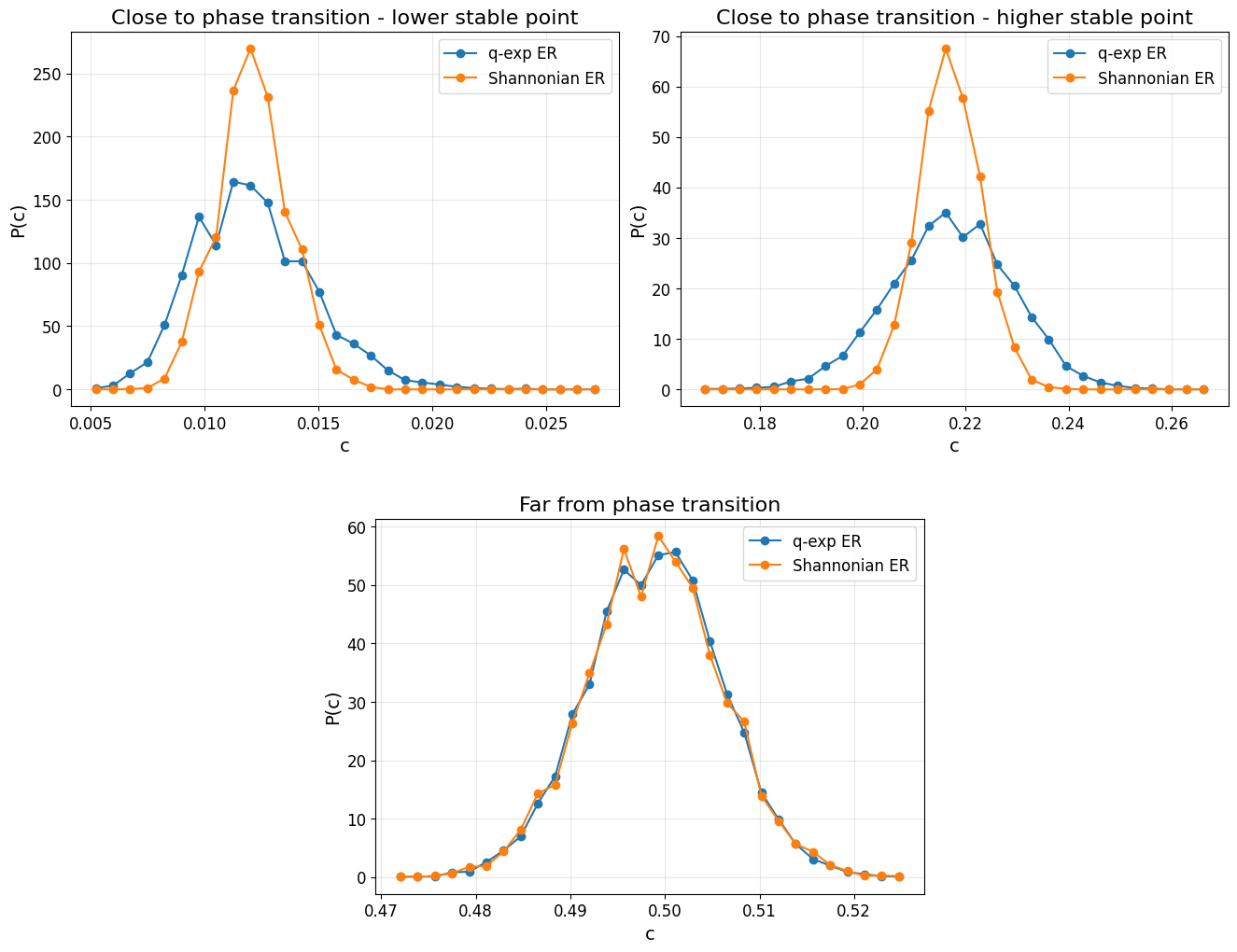}
    \caption{Comparison of distributions link densities for two situations: close to the phase transition (top two graphs) with $r=2.7$ and $\theta=5.154$ and far from it (bottom graph) with $r=1000$ and $\theta=1$. We also plot the corresponding Erd\H{o}s--R\'enyi ensembles (fitted to reproduce the same average link density). For the top two graphs, we used two different initial conditions and converged to two different fixed points of the saddle-point equation. Used ensembles contained 10000 graphs with 100 nodes.}
    \label{fig:link_density_distributions}
\end{figure*}

\begin{figure*}[!t]
    \centering
    \includegraphics[width=0.63\textwidth]{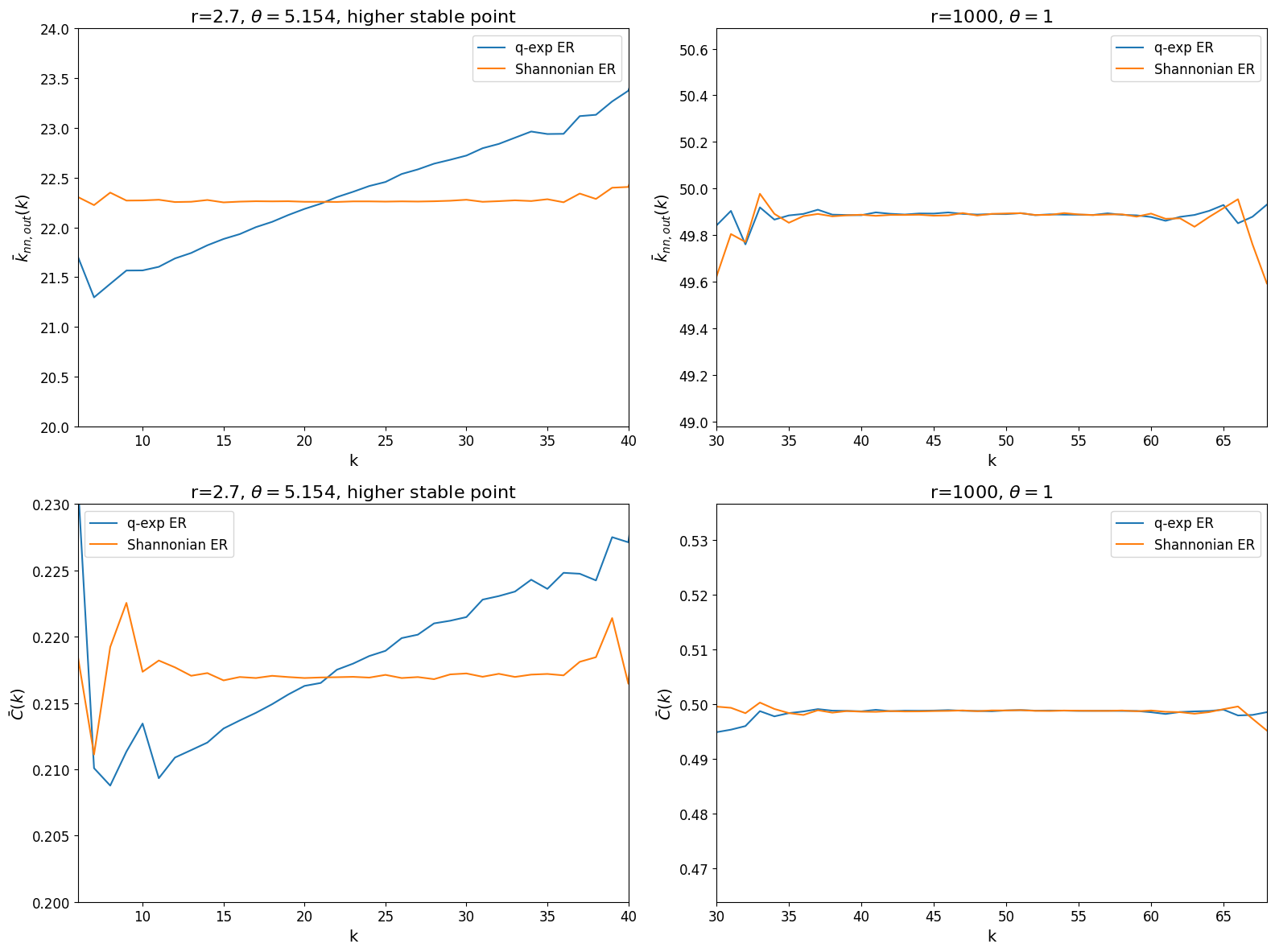}
    \caption{Comparison of the ANND and clustering coefficient depending on node degree for the non-Shannon ensembles close to the phase transition (left) and far from it (right). Interestingly, near the phase transition, the $q$-exponential model develops positive assortativity and increasing clustering with degree. For extreme values of degrees, there are less available data and therefore the computation is more noisy. Here we have used the same ensembles as for Fig.~\ref{fig:link_density_distributions}.}
    \label{fig:annd_clustering}
\end{figure*}

Having established the theoretical properties of the $q$-ER model defined by Eq.~\eqref{IV.30.kl}, we now turn to numerical experiments, both to validate our analytical predictions and to study further network properties such as clustering and average nearest-neighbor degree. Graph ensembles are sampled using the Metropolis--Hastings algorithm with a symmetric link-flip proposal (see Appendix~\ref{sec:metropolis_hastings}), for which the acceptance probability takes the form
\begin{eqnarray}
\alpha &=& \min\!\left(1, \frac{P_q(G')}{P_q(G_t)}\right) \nonumber \\[2mm]
&=& \min\!\left(1, \frac{\exp_{2-q}\!\left(-\theta L(G')\right)}{\exp_{2-q}\!\left(-\theta L(G_t)\right)}\right),
\end{eqnarray}
where the partition function cancels in the ratio.

\subsubsection{Link density and percolation}
We considered two complementary methods for approximating the link density. The first one is a perturbative approach based on Eq.~\eqref{eq:er_marginal_link_probability}, while the second one involved solving the saddle-point equation~\eqref{eq:saddle_point_equation}. Fig.~\ref{fig:link_densities} compares the predictions of these approaches with numerical simulations. For 10 values of $(1-q)$ evenly spaced between $10^{-6}$ and $10^{-3}$, and for $\theta \in \{1,3,5\}$, we generated ensembles of $1000$ graphs with $100$ nodes, computed their link densities, and compared them with the theoretical predictions. The results indicate that the linear perturbative approximation loses accuracy rapidly, whereas the saddle-point solution remains in excellent agreement with the numerical data. Furthermore, as $q$ deviates further from unity, the link density for all values of $\theta$ approaches $0.5$, consistent with our analytical predictions.  

Let us next examine, how the link densities are distributed among the ensemble. Two representative regimes are considered: one close to the phase transition and one far from it. We focus on the case $r=(1-q)N=2.7$, for which a phase transition was identified at $\theta_{\mathrm{crit}}=5.154$, with the corresponding saddle-point solutions $c_{\mathrm{low}}=0.0118$ and $c_{\mathrm{high}}=0.2177$. To probe the resulting distributions, we generate two ensembles, each with a different initial condition for sampling: one starting with a graph with density $c_{\mathrm{init}}=0.01$ and the other with $c_{\mathrm{init}}=0.3$. For each ensemble, $10^4$ graphs are generated, and the resulting link-density histograms are computed. For comparison, we also generate an ensemble with $r=1000$ and $\theta=1$, corresponding to a regime far from the transition. Finally, we construct Shannon ensembles tuned to reproduce the same average link densities as the corresponding non-Shannon ensembles.

The results are summarized in Fig.~\ref{fig:link_density_distributions}. The two distinct initial conditions converge to different fixed points of the saddle-point equation, as we can see from the corresponding link-density distributions. The measured mean values, $\langle c \rangle_{\mathrm{low}} = 0.0120$ and $\langle c \rangle_{\mathrm{high}} = 0.2173$, are in excellent agreement with the theoretical predictions. Notably, in the vicinity of the phase transition, the non-Shannon ensembles exhibit substantially broader distributions than their Shannon counterparts. The corresponding standard deviations are $\sigma_{\mathrm{low}} = 0.0018$ and $\sigma_{\mathrm{high}} = 0.0082$ for the non-Shannon ensembles, compared to $\sigma_{\mathrm{low}} = 0.0011$ and $\sigma_{\mathrm{high}} = 0.0041$ for the Shannon ensembles.

By contrast, for the ensemble far from the critical point, the distributions are nearly indistinguishable, with $\langle c \rangle = 0.4995$ and $\sigma = 0.005$ in both cases. This behavior confirms our analytical expectation that, for $q$ sufficiently far from unity, the model reduces to the Erd\H{o}s--R\'enyi random graph with link probability $1/2$, whereas for $q$ close to unity the two ensembles display markedly different statistical properties.

We also investigated the percolation transition, which is a characteristic phenomenon of the ER model. The overall behavior appears to be the same for the $q$-ER, the only noticeable difference being an increased variance in the size of the giant connected component. Further results are provided in Appendix~\ref{sec:qERM_clustering_percolation}.

\subsubsection{Average nearest-neighbor degree and clustering}

Let us now compare additional structural properties of the networks. We find no significant differences in the degree distributions between the $q$-exponential and exponential (Shannon) ensembles. In contrast, the average nearest-neighbor degree (ANND) and the clustering coefficient exhibit notable differences between the q-exponential model near the phase transition and the regime far from criticality. These results are shown in Fig.~\ref{fig:annd_clustering}, where we analyze the same ensembles as in Fig.~\ref{fig:link_density_distributions} and evaluate the degree dependence of the ANND and clustering coefficient.\footnote{For each graph, the ANND and clustering coefficient are first computed at the node level. Nodes with the same degree are then grouped across the ensemble, and the corresponding quantities are averaged within each group.}

For the ANND, the non-Shannon model near the transition displays positive assortativity, accompanied by an increase of the clustering coefficient with node degree. We attribute this behavior to the presence of positive edge-to-edge correlations, in agreement with the analytical prediction in Eq.~\eqref{eq:er_edge_covariance}. By contrast, in the regime $r=1000$ and $\theta=1$, the model approaches the Shannon limit, and both assortativity and clustering remain approximately constant, as expected for the Erd\H{o}s--R\'enyi model. 

\begin{figure}[!t]
    \centering
    \includegraphics[width=0.48\textwidth]{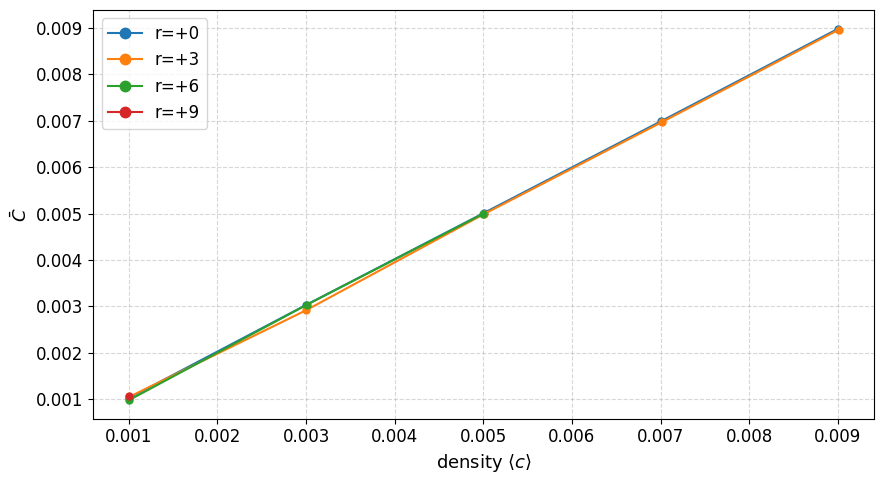}
    \caption{Average clustering coefficient as a function of graph density for the $q$-ER model, for target average degrees $\langle k \rangle \in \{1, 3, 5, 7, 9\}$ and $r \in \{0, 3, 6, 9\}$, based on ensembles of 1000 graphs with $n=1000$ nodes. No significant deviation from the Shannon ($r=0$) case is observed, in contrast to the $q$-exponential CM (cf.\ Fig.~\ref{fig:conf_model_clustering_vs_density}).}
    \label{fig:er_clustering_vs_density}
\end{figure}

We also studied the overall average clustering coefficient as a function of graph density for varying $r$, shown in Fig.~\ref{fig:er_clustering_vs_density}. Across all target degrees $\langle k \rangle \in \{1, 3, 5, 7, 9\}$ and values of $r \in \{0, 3, 6, 9\}$, the clustering remains essentially indistinguishable from the Shannon ($r=0$) baseline. We attribute this to the nature of the higher-order interactions in the $q$-ER model: as seen in Eq.~\eqref{eq:effective_hamiltonian_erm}, the corrections couple all pairs of links equally through powers of the total link count $L(G)$, with no separate parameter to enforce a preference for triangle formation. This is in sharp contrast to what we will show in the next Section for the $q$-exponential CM, where varying $r$ produces substantial deviations from the Shannon case and has a remarkable impact on the clustering coefficient.

\section{$q$-exponential Configuration Model}
\label{sec:tsallis_park_newman}
We finally turn to the $q$-exponential  generalization of the CM introduced in Sec.~\ref{sec:shannon_CM} (which we denote as the $q$-CM), which has constraints on the degrees of all nodes. The corresponding Hamiltonian is still given by Eq.~\eqref{eq:park_newman_hamiltonian}, while the resulting $q$-exponential graph probability is given by
\begin{equation}
    P_q(G) \;=\; \frac{1}{\mathcal{Z}_{q}(\boldsymbol{\theta})}
    \exp_{2-q}\!\left(-\sum_{i=1}^n\sum_{j < i} (\theta_i + \theta_j) a_{ij}\right),
\end{equation}
with partition function
\begin{eqnarray}
\mathcal{Z}_{q}(\boldsymbol{\theta}) = \ \sum_{\{a_{ij}\}} \exp_{2-q}\left(-\sum_{i=1}^n\sum_{j < i}(\theta_i + \theta_j)a_{ij}\right).
\end{eqnarray}
For given $q$, the conditions for enforcing a target degree sequence $\{k^*_i\}_{i=1}^n$ are analogous to those in Eq.~\eqref{eq_parametersCM}:
\begin{equation}
k_i^*\equiv \langle k_i\rangle =  \sum_{j\neq i}\langle a_{ij}\rangle=  \sum_{j\neq i}p_{ij},\quad i=1,n,
\label{eq_parametersqCM}
\end{equation}
where here $p_{ij}=\langle a_{ij} \rangle$ represents the \emph{marginal} connection probability and, however, the overall $P_q(G)$ does not factorize into a product of terms involving $p_{ij}$, because of overall dependencies between all pairs of nodes, as for the $q$-ER model.  

We can compute $\langle a_{kl} \rangle$ for the $q$-CM by summing out all degrees of freedom except link $(k,l)$ (see Appendix~\ref{sec:avg_link_density_cm} for the derivation). Introducing the reduced partition function and reparametrized multipliers
\begin{eqnarray*}
    \tilde{\mathcal{Z}}^{(k,l)}_{q}(\boldsymbol{\theta}) &\equiv&\!\!\!\! \sum_{\substack{\{a_{ij}\}\\(i,j)\neq (k,l)}} \exp_{2-q}\!\left(- \sum_{\substack{i<j\\(i,j)\neq (k,l)}}
    (\theta_i + \theta_j) a_{ij}\right), \\[2mm]
    \tilde{\theta}_i^{(k,l)} &\equiv& \frac{\theta_i}{1-(q-1)(\theta_k + \theta_l)}\, ,
\end{eqnarray*}
we obtain
\begin{equation}
    \langle a_{kl} \rangle \ = \ \exp_{2-q}\left(-(\theta_k + \theta_l)\right)\frac{\tilde{\mathcal{Z}}^{(k,l)}_{q}(\boldsymbol{\tilde{\theta}}^{(k,l)})}{\mathcal{Z}_{q}(\boldsymbol{\theta})}\, .
    \label{eq:avg_link_density_park_newman}
\end{equation}
This expression is reminiscent of Eq.~\eqref{eq:avg_link_density} obtained in the $q$-ER model. Although it would be desirable to further simplify the numerator,
${\tilde{\mathcal{Z}}^{(k,l)}_{q}(\boldsymbol{\tilde{\theta}}^{(k,l)})}$,
the $q$-exponential of a sum does not factorize in a simple manner, and Eq.~\eqref{eq:avg_link_density_park_newman} is likely the most compact form attainable.
In principle, we would again carry out a perturbative analysis, analogous to the one performed in Sec.~\ref{sec:Erdos_Renyi_moments_and_cumulants}. In what follows, however, we focus primarily on the behavior of network properties and evaluate them numerically.

\subsection{Network properties}
\label{sec:conf_model_network_properties}
%
\label{sec:tsallis_park_newman_network_properties}
Among the virtually infinite possible choices for generating target degree sequences, we consider the convenient strategy indicated in Eqs.~\eqref{eq_FM} and~\eqref{eq_parametersFM}: we sample $\{s_i\}_{i=1}^n$ i.i.d. from a given distribution, while adjusting $z$ to enforce a desired overall link density $c^*$. 
To pick a realistic distribution, we refer to a study of interbank networks~\cite{Gabrielli2024} where the empirical distribution of node `strengths' $\{s_i\}_{i=1}^n$ was found to be well approximated by a log-normal distribution with $\sigma=2.28$, $\mu=-\sigma^2/2$. Since the relationship to the parameters $\theta_i$ is given by $x_i = e^{-\theta_i}$, we can sample $\theta_i$ from a normal distribution with $\sigma=2.28$ and $\mu=\sigma^2/2$. Inserting these values into the $p_{ij}$ in Eq.~\eqref{eq_FM} and summing over all pairs as in Eq.~\eqref{eq_parametersFM}, we obtain the implied link density in the reference ($q=1$) Shannonian CM and we adjust $z$ to enforce a desired value $c^*$ that we consider as the target one. We then consider different values of $q$ (or $r$) and recalculate $z$ for each of them, to match the same target density $c^*$ for the general $q$-CM, while keeping strengths $\{s_i\}_{i=1}^n$ the same. 
With this strategy we can vary $z$ to explore the space of densities, and compare the CM and $q$-CM for the same target level of density. 
Since the parameter $z$ is multiplicative, in terms of $\theta_i$ we have to find an additive shift denoted by $\theta_{\rm{shift}}$, which will recover the target density of the network.

Let us now see whether the $q$-exponential model gives significant corrections in terms of network properties and whether we will also observe signs of a phase transition. To this end, we choose target average node degrees $k^* \in \{1, 3, 5, 7, 9\}$ (which corresponds to choosing target average graph density $c^* = k^*/(n-1)$. For each of them, we choose several values of $r=(1-q)N$ ($N$ is the maximum number of links, $N=n(n-1)/2$) and for each we find the corresponding $\theta_{\rm{shift}}$ to recover the target density\footnote{We find the corresponding $\theta_{\rm{shift}}$ by bisection, i.e. we choose an interval of possible values of $\theta_{\rm{shift}}$ and iteratively narrow it down until we reach given precision.}. Importantly, this time we try both $r>0$ (corresponding to $q<1$) and $r<0$ (corresponding to $q>1$). We then sample 1000 networks with  $n=1000$ nodes using the Metropolis--Hastings algorithm, similarly to the $q$-ER model (Sec.~\ref{sec:qERM_network_properties}), with a symmetric link-flip proposal (see Appendix~\ref{sec:metropolis_hastings}). The acceptance probability is given by
\begin{eqnarray}
    \alpha \ &=& \ \min\left(1, \frac{P_q(G')}{P_q(G_t)}\right) \nonumber \\[2mm] &=& \ \min\left(1, \frac{\exp_{2-q}\left(-\sum_{i=1}^n\theta_i k_i(G')\right)}{\exp_{2-q}\left(-\sum_{i=1}^n\theta_i k_i(G_t)\right)}\right).
\end{eqnarray}
%

\begin{figure}[t]
    \centering
    \includegraphics[width=0.48\textwidth]{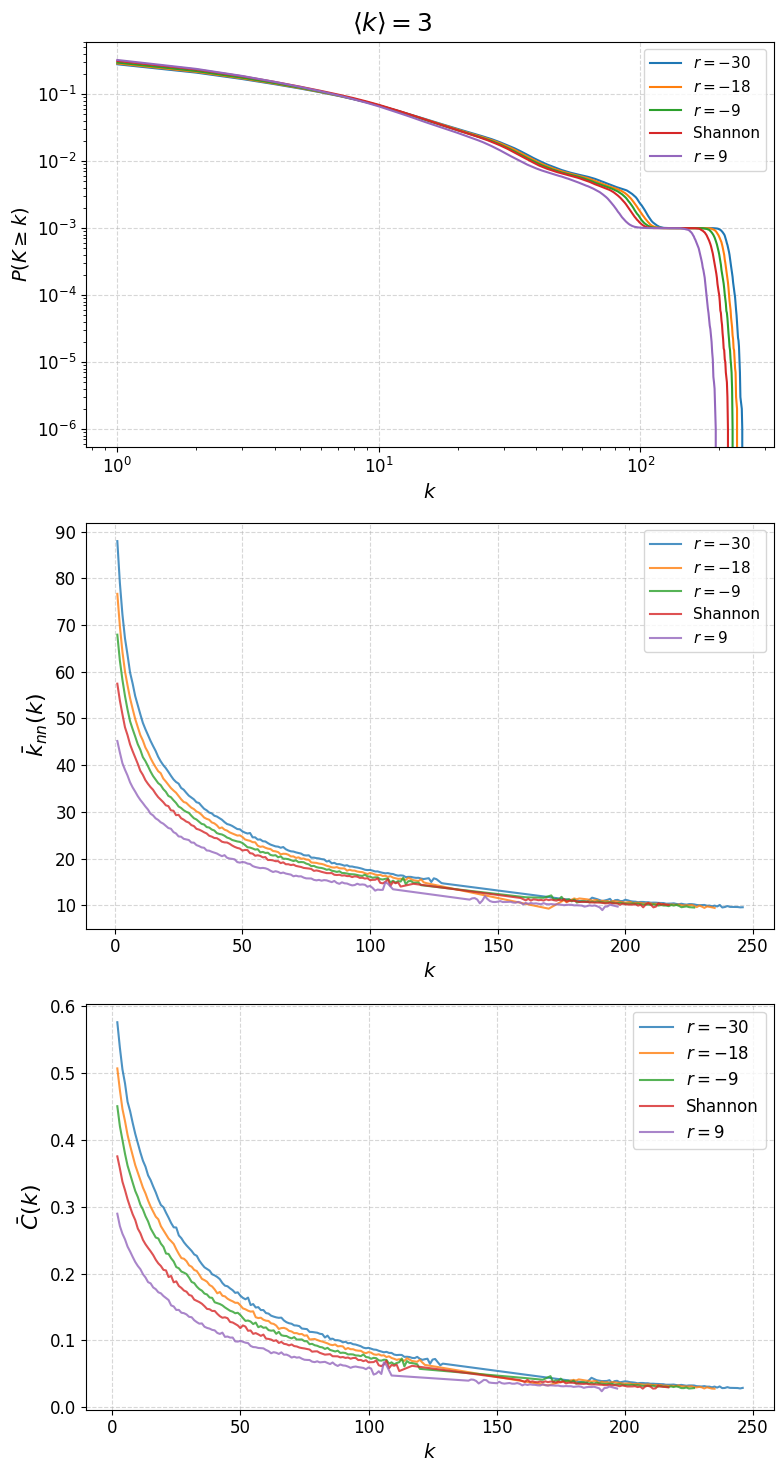}
    \caption{Degree-resolved network properties of the $q$-exponential CM for target average degree $k^* = 3$ and $n=1000$ nodes, for $r \in \{-30, -18, -9, 0, 9\}$. \textit{Top}: complementary cumulative degree distribution. \textit{Middle}: average nearest-neighbor degree $\bar{k}_{nn}(k)$. \textit{Bottom}: clustering coefficient $\bar{C}(k)$. While the degree distributions are nearly identical across $r$ values, both the ANND and clustering are strongly modulated by $r$, with more negative $r$ producing higher correlations and clustering at low degrees. Each curve is averaged over an ensemble of 1000 graphs.}
    \label{fig:conf_model_network_properties_k3}
\end{figure}

We find that, similar to the $q$-ER model, certain target densities are irreconstructible, signifying a similar type of phase transition. For the values of $r$ for which reconstruction was possible, however, we observe rich degree-resolved structure. Fig.~\ref{fig:conf_model_network_properties_k3} shows the complementary cumulative degree distribution (CCDF), average nearest-neighbor degree $\bar{k}_{nn}(k)$, and clustering coefficient $\bar{C}(k)$ for target average node degree $k^* = 3$ across a range of $r$ values. 
A full empirical study across other target degrees is presented in Appendix~\ref{sec:conf_model_empirical_results}.
The degree distributions are nearly indistinguishable across $r$, reflecting the fact that the $\theta_i$ distribution is held fixed and only the additive shift $\theta_{\mathrm{shift}}$ is tuned to recover the target density. In contrast, both the ANND and the clustering coefficient are strongly sensitive to $r$: more negative values of $r$ (corresponding to $q > 1$) produce markedly higher ANND and clustering at low degrees, while positive $r$ suppresses them.
We therefore find tunable clustering and assortativity profiles, similar to those reported in studies where the degree sequence is kept fixed and extra constraints on the clustering coefficients are introduced~\cite{serrano2005tuning}. However, here the only actual constraint is the degree sequence alone as in the traditional CM, and the tunable clustering profiles are automatically produced by values $q\ne 1$.

\begin{figure}[t]
    \centering
    \includegraphics[width=0.5\textwidth]{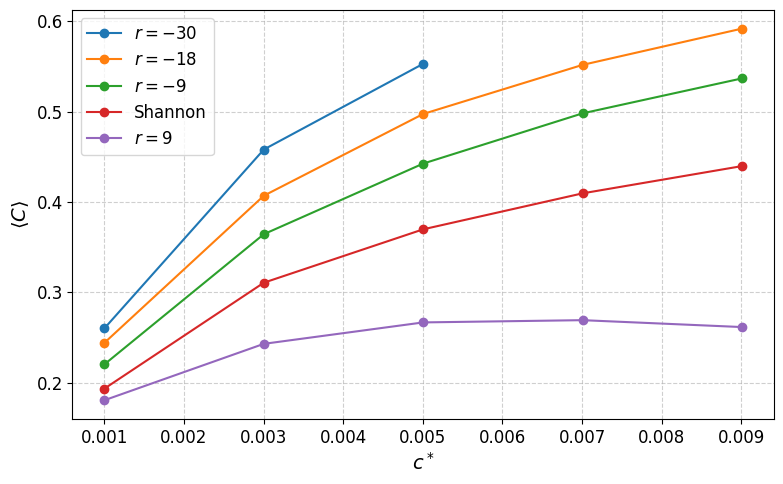}
    \caption{Average clustering coefficient as a function of target network density $c^*$ for the $q$-exponential CM, for target average node degrees $k^* \in \{1, 3, 5, 7, 9\}$. Each band of curves corresponds to a fixed target density with varying $r$, illustrating that the parameter $r$ independently controls the clustering coefficient while leaving the degree sequence unchanged. The accessible range of clustering grows substantially with increasing target density. Each data point is averaged over an ensemble of 1000 graphs on $n=1000$ nodes.}
    \label{fig:conf_model_clustering_vs_density}
\end{figure}

A striking consequence is that the parameter $r$ acts as an independent knob for the average local clustering coefficient \emph{at fixed target density}, in sharp contrast to the standard (Shannon) CM where the degree sequence alone determines the clustering. Indeed, as shown in Fig.~\ref{fig:conf_model_clustering_vs_density}, there is a broad range of accessible values of the average clustering coefficient for a given target density, demonstrating that the $q$-exponential CM provides a genuinely richer structural repertoire than its Shannonian counterpart. Remarkably, this allows for the coexistence of sparsity and clustering, which is a widespread empirical property of real networks that is impossible to replicate in the usual $q=1$ case, for both the ER and the CM. 
Moreover, comparing Fig.~\ref{fig:conf_model_clustering_vs_density} with Fig.~\ref{fig:er_clustering_vs_density}, we stress that the dependence, for fixed density, of the average clustering coefficient on the value of $q$ is found only for the $q$-CM and not for the $q$-ER model. 
This means that the mechanism required to produce a non-trivial clustering is not the interaction among links \emph{per se}, but the \emph{combination of interaction and heterogeneity}. 
Indeed, as already recalled in Sec.~\ref{sec:relation}, also in traditional homogeneous ERGs such as the edge-triangle model~\cite{strauss1986general,park2005solution,coolen2017generating} it is not possible to obtain simultaneously a large clustering and a low density (and more generally any combination of values for the two quantities), despite the presence of interactions and the existence of two separate Lagrange multipliers designed to tune the two quantities.
On the other hand, heterogeneity alone (as in the ordinary CM) is not enough to produce a large clustering. 
Therefore the combination of the two ingredients seems necessary.

\section{Conclusions}
\label{sec:conclusions}
In this paper, we have introduced the most agnostic inferential framework for network ensembles with given expected properties, based on the maximization of generalized (Uffink-type) entropies that extend beyond Shannon entropy. 
The generalization is required in light of the recent (re)discovery~\cite{Uffink1995,Jizba2019,Jizba2020} that the axiomatic definition of a proper inferential MEP unavoidably leads to a class of entropies that is more general than Shannon's, unlike initially claimed~\cite{Shore1980,Shore1981}.
The corresponding maximum-entropy probability distribution is $q$-exponential, with $q=1$ reducing to the traditional Shannon case. 
The $q$-exponential turns the entire class of ERGs, which include the most popular network models, into $q$-ERGs, which we have investigated here for the first time. 
The resulting framework provides a promising link between the simplicity of maximum-entropy inference (within the MEP approach) and the complexity of real-world network topology. Within this setting, the structure of the underlying Uffink MEP functionals admits a clear operational meaning: the function $f$ determines the scaling behavior of the entropy for uniform distributions (which is relevant for classification purposes~\cite{Thurner,Thurner:2018}), while the positive parameter $q$ quantifies correlations in the system through the associated MEP distributions.\\

Our analysis of the simplest $q$-ERGs (defined by enforcing the overall link density and the degree sequence respectively) yielded several key insights:

\begin{itemize}
    \item {\bf Non-independence of links:} Unlike the classical Shannonian model, where links are independent Bernoulli variables, the $q$-exponential framework couples all potential links through the non-additive nature of the entropy. This can be efficiently captured via the notion of effective Hamiltonian, which features extra effective properties that will appear in the network unavoidably as $q\ne 1$, but without dedicated extra Lagrange multipliers tuning them: the extra features are all driven by the value of $q$, with no need to make arbitrary choices of extra constraints to be added to the original Hamiltonian. We derived perturbative expressions for covariances, showing that for $q \ne 1$, the model inherently induces nonzero correlations between links. 
    \item {\bf Higher-order networks from simple constraints:} 
    The $q$-exponential setting provides a natural way of creating higher-order link dependencies from the `usual' simple constraints (link density, degree sequence) without introducing such dependencies in an \emph{ad hoc} manner via the addition of arbitrary higher-order constraints (such as the counts of two-stars, triangles, etc.). This result is parsimonious and particularly relevant in light of the current discussion around higher-order networks and hypergraphs~\cite{battiston2022higher,bick2023higher,bianconi2021higher,boccaletti2023structure,peixoto2026graphs}. We are therefore confronted with the fascinating result that, in absence of prior information ruling out values $q\ne 1$, the most agnostic maximum-entropy ensemble of graphs with given number of links, or given degree sequence, does not have independent edges and is therefore a higher-order network model, without the need to input additional (and arbitrarily chosen) higher-order constraints. 

    \item {\bf The critical scaling regime:} We identified that the `Shannon degeneracy' (the tendency of $q$-exponential statistics to collapse into exponential Boltzmann--Gibbs statistics in the large-$N$ limit) can be overcome by the specific scaling $(1-q)N = r$ with $r$ fixed and finite. This reveals that for the $q$-exponential paradigm to remain relevant for large networks, the non-extensivity parameter $1-q$ must scale inversely with the system size.
    \item {\bf Phase transitions and structural complexity:} Through a saddle-point analysis, in the $q$-ER model we proved the existence of a phase transition between sparse and dense network regimes. Crucially, we observed that near this criticality, the ensemble exhibits structural features --- such as degree-dependent clustering and assortativity --- that are absent in the standard Erd\H{o}s--R\'{e}nyi model. The behavior of the $q$-ER is in fact typical of standard homogeneous ERGs with the addition of higher-order constraints~\cite{strauss1986general,park2004solution,annibale2015two,park2005solution,coolen2017generating}, which are however unnecessary here.
    \item {\bf Generalized Configuration Models:} 
    Our extension of the CM to the $q$-CM demonstrates that the $q$-exponential framework is sufficiently robust to accommodate degree constraints. We showed that the parameter $r$ enables the tuning of higher-order structural properties, such as the average nearest-neighbor degree and the local clustering coefficient, while keeping the degree sequence constrained, thereby providing an additional degree of freedom for network reconstruction. The resulting tunable clustering and assortativity profiles are similar to those obtained in more complicated models~\cite{serrano2005tuning} requiring the introduction of extra constraints on the local clustering coefficients. Remarkably, for appropriate values of $q$ the average local clustering coefficient can be made large even for low link density, a feature that is widespread in real networks but impossible to replicate via the ordinary CM, via the $q$-ER, or even via traditional homogeneous ERGs with edge-triangle constraints. This indicates the need for a combination of link dependencies and heterogeneity in order to replicate sparse clustered graphs via maximum-entropy models.

\end{itemize}


In future work, we plan to extend this framework to other class of graphs such as multiplex and weighted networks, where the non-additive nature of the entropy may additionally capture inter-layer correlations and the heavy-tailed nature of various distributions commonly observed in empirical data. Overall, our results indicate that the `physics of information' in networks is highly sensitive to the choice of entropic functional, and that non-Shannon entropies offer a mathematically rigorous and phenomenologically viable approach for modeling the intrinsic dependencies within complex systems.

\begin{acknowledgments}
%
We thank Constantino Tsallis and Andrea Somazzi for useful discussions and comments on the manuscript.  P.J. was supported by the Czech Science Foundation Grant (GA\v{C}R), Grant No.
25-18105S. 
This publication is part of the project ``Redefining renormalization for complex networks’’ with file number OCENW.M.24.039 of the research programme Open Competition Domain Science Package 24-1, which is (partly) financed by the Dutch Research Council (NWO) under the grant \url{https://doi.org/10.61686/PBSEC42210}.

\end{acknowledgments}

\clearpage
\appendix

\begin{widetext}
\section{Derivation of the $q$-exponential ERG distribution}
\label{sec:qERG_derivation}
%
We derive the general $q$-exponential graph probability stated in Sec.~\ref{sec:qERGs}, starting from the Lagrange functional $\mathcal{L}_q[P]$ in Eq.~\eqref{eq:lagrange_functional}. Computing the partial derivative with respect to $P_q(G')$ and demanding it to vanish for all $G' \in \mathcal{G}_n$, we find
\begin{eqnarray}
    \frac{\partial}{\partial P_q(G')}\mathcal{L}_q[P] \ &=& \  \frac{q}{1-q} P_q^{q-1}(G') - \sum_{i=1}^{K}\theta_i c_i(G') - \alpha \ = \ 0\qquad \forall G' \in \mathcal{G}_n\, .
    \label{17.bh}
\end{eqnarray}
To eliminate $\alpha$, we multiply~\eqref{17.bh} by $P_q(G')$ and sum over all $G' \in \mathcal{G}_n$, giving
\begin{eqnarray}
    \alpha \ = \  \frac{q}{1-q} \sum_{G \in \mathcal{G}_n} P_q^q(G) \ - \  \sum_{i=1}^{K}\theta_i \langle c_i \rangle\, .
    \label{eq_alpha}
\end{eqnarray}
Substituting back Eq.~\eqref{eq_alpha} into Eq.~\eqref{17.bh} and multiplying by $(1-q)/q$ yields
\begin{eqnarray}
    &&P_q^{q-1}(G') \ = \ \sum_{G \in \mathcal{G}_n} P_q^q(G) \ + \ \frac{1-q}{q} \sum_{i=1}^{K}\theta_i (c_i(G') - \langle c_i \rangle)\, ,
\end{eqnarray}
which can be conveniently rewritten as
\begin{eqnarray}
    &&P_q(G') \ = \ \left(\sum_{G \in \mathcal{G}_n} P_q^q(G)\right)^{1/(q-1)} \left[ 1 - \frac{q-1}{q}\frac{1}{\sum_{G \in \mathcal{G}_n} P_q^q(G)}\left(\sum_{i=1}^{K}\theta_i (c_i(G') - \langle c_i \rangle)\right) \right]_+^{1/(q-1)}.
\end{eqnarray}
The prefactor $\left(\sum_{G \in \mathcal{G}_n} P_q^q(G)\right)^{1/(q-1)}$ can be identified with both the inverse partition function $\mathcal{Z}^{-1}_q({\bm{\theta}})$ and the inverse Uffink entropy $U^{-1}_q[P_q]$ of $P_q$ itself -- which is an indication of self-referentiality:
\begin{eqnarray}
    \mathcal{Z}_q({\bm{\theta}}) \ \equiv \ \left(\sum_{G \in \mathcal{G}_n} P_q^q(G)\right)^{1/(1-q)}\ =\ U_q[P_q]\, .
    \label{eq:partition_fn}
\end{eqnarray}
Now, exploiting the duality relation 
\begin{equation}
\exp_q(-x) \ = \  [1 \ - \  (1 - q)\ \! x]^{1/(1-q)}_+= \  [1 \ + \  (q - 1)\ \! x]^{-1/(q-1)}_+ \ = \ \left[\exp_{2-q}(x)\right]^{-1},  \quad (q\in \mathbb{R})\, ,
\label{eq_duality}
\end{equation}
mapping the $q$-exponential of $-x$ to the inverse $(2-q)$-exponential of $x$, the general $q$-exponential graph probability $P_q(G)$ can be written as
\begin{eqnarray}
        P_q(G) & =&  \frac{1}{\mathcal{Z}_q({\bm{\theta}})} \left[ 1 - \frac{q-1}{q}\frac{1}{\mathcal{Z}_q^{1-q}({\bm{\theta}})}\left(\sum_{i=1}^{K}\theta_i (c_i(G) - \langle c_i \rangle)\right) \right]_+^{1/(q-1)} \nonumber\\[2mm]
        & =&  \frac{1}{\mathcal{Z}_q({\bm{\theta}})} \left[ \exp_q\left(\frac{1}{q\mathcal{Z}_q^{1-q}({\bm{\theta}})}\sum_{i=1}^{K}\theta_i (c_i(G) - \langle c_i \rangle)\right) \right]^{-1}\\[2mm]
        &=& \frac{1}{\mathcal{Z}_q({\bm{\theta}})} \exp_{2-q}\!\left(-\frac{1}{q\mathcal{Z}_q^{1-q}({\bm{\theta}})}\sum_{i=1}^{K}\theta_i (c_i(G) - \langle c_i \rangle)\right).
\label{eq:tsallis_model_general_appendix}
\end{eqnarray}
To resolve the self-referential dependence on both $\mathcal{Z}_q=U_q[P_q]$ and $\langle c_i\rangle$, we use the $q$-exponential addition formula (cf.~\cite{Tsallis2009})
\begin{eqnarray}
    \exp_{2-q}(x+y) 
   \ = \ \exp_{2-q}(x)\exp_{2-q}\!\left(\frac{y}{1+(q-1)x}\right).
    \label{eq:q_exp_sum_formula}
\end{eqnarray}
With this we can write
\begin{eqnarray}
       &&\mbox{\hspace{-5mm}} \exp_{2-q} \Bigl(-\frac{1}{q\mathcal{Z}_q^{1-q}}\sum_{i=1}^{K}\theta_i (c_i(G) - \langle c_i \rangle)\Bigr) \nonumber \\[2mm] &&= \ \exp_{2-q}\!\left(\frac{1}{q\mathcal{Z}_q^{1-q}}\sum_{i=1}^{K}\theta_i \langle c_i\rangle \right)\exp_{2-q}\!\left(-\frac{1}{q\mathcal{Z}_q^{1-q} + (q-1)\sum_{i=1}^{K}\theta_i \langle c_i \rangle}\sum_{i=1}^{K}\theta_i c_i(G)\right).
\end{eqnarray}
The first factor on the right-hand side is independent of $G$ and therefore cancels in both numerator and denominator of Eq.~\eqref{eq:tsallis_model_general_appendix}. We can thus define reparametrized multipliers
\begin{equation}
    \tilde{\theta}_i \ \equiv \ \frac{\theta_i}{q\mathcal{Z}_q^{1-q} + (q-1)\sum_{i=1}^{K}\theta_i \langle c_i \rangle}\, ,
\end{equation}
and obtain
\begin{equation}
    P_q(G) \ = \ \frac{1}{\mathcal{Z}_q} \exp_{2-q}\!\left(-\sum_{i=1}^{K}\tilde{\theta}_i c_i(G)\right).
    \label{eq:q_exp_prob_reparametrized}
\end{equation}
As shown in Ref.~\cite{Jizba2017}, the parameters $\tilde{\theta}_i$ can be identified with Casimir invariants associated with the group of reparametrizations of Lagrange multipliers that preserve form invariance of the distribution (up to an arbitrary shift in $c_i$'s). Dropping the tilde and writing $H(G,{\bm{\theta}}) = \sum_{i=1}^{K}\theta_i c_i(G)$, Eq.~\eqref{eq:q_exp_prob_reparametrized} takes the compact form~\eqref{eq:prob_with_hamiltonian}.

\section{Large $N$ behavior - detailed computations of the $q$-exponential Erd\H{o}s--R\'enyi model}
\label{sec:appendix_qER_largeN}

We derive the two results stated in Sec.~\ref{sec:qER_largeN}: the variational formula for the free entropy density~\eqref{eq:free_entropy_limit} and the large-deviation form of the link-density probability.

\subsection*{Step 1: Bounds on the probability}

Since all $\binom{N}{k}$ adjacency matrices with exactly $k$ links are equally weighted by $\exp_{2-q}(-\theta k)$, the probability of observing link density $c = k/N$ is
\begin{equation}
    P\!\left(\frac{L(G)}{N} = c\right)
    \ = \  \binom{N}{Nc} \frac{\exp_{2-q}(-\theta Nc)}{\mathcal{Z}_{q,N}(\theta)}\, .
    \label{appendix_IV.33.cb}
\end{equation}
We can use standard information-theoretic bounds~\cite{Cover2006}
\begin{equation}
    \frac{1}{N+1}\,e^{N H_b(c)} \ \leq \ \binom{N}{Nc} \ \leq \ e^{N H_b(c)}\, ,
\end{equation}
where $H_b(c) = -c\ln c - (1-c)\ln(1-c)$ is the binary entropy. Then let us rewrite the $q$-exponential as a standard exponential via
\begin{eqnarray}
    e^{N H_b(c)}\,\exp_{2-q}(-\theta Nc)
    \ \equiv \ e^{N\phi_q(c,\theta,N)},
\end{eqnarray}
with the \emph{effective free entropy density} [Eq.~\eqref{eq:effective_free_entropy} of the main text]

\begin{eqnarray}
    \phi_q(c,\theta, N)
    \ = \  H_b(c) \ - \  \frac{1}{N(1-q)}\ln\!\left[1 \ + \ (1-q)\theta Nc\right].
\label{eq:appendix_effective_free_entropy}
\end{eqnarray}
Using this rewritten form, the bounds become
\begin{eqnarray}
    \frac{e^{N\phi_q(c,\theta,N)}}{(N+1)\,\mathcal{Z}_{q,N}(\theta)}
    &\leq& P\!\left(\frac{L(G)}{N}=c\right)
    \ \leq  \ \frac{e^{N\phi_q(c,\theta,N)}}{\mathcal{Z}_{q,N}(\theta)}\, .
    \label{eq:appendix_prob_bounds}
\end{eqnarray}

\subsection*{Step 2: Variational formula for $\Phi_q$}

Let $c_{\max} \equiv \arg\max_c \phi_q(c,\theta,N)$. We examine each bound in~\eqref{eq:appendix_prob_bounds} separately.
\begin{itemize}
    \item \textbf{Lower bound.} Since $P\!\left(\frac{L(G)}{N}=c\right)\leq 1$, the lower bound in~\eqref{eq:appendix_prob_bounds} gives $\frac{1}{N+1}e^{N\phi_q(c,\theta,N)}\leq\mathcal{Z}_{q,N}(\theta)$. This holds in particular for $c=c_{\max}$; taking logarithms and dividing by $N$,
    \begin{equation}
        -\frac{\ln(N+1)}{N}  \ + \ \phi_q(c_{\max},\theta,N) \ \leq \ \frac{\ln\mathcal{Z}_{q,N}(\theta)}{N}\, .
        \label{eq:appendix_Z_lower_bound}
    \end{equation}

    \item \textbf{Upper bound.} Summing the upper bound of~\eqref{eq:appendix_prob_bounds} over all $N+1$ possible values of $c$ (one for each $k\in\{0,\ldots,N\}$) and using $\sum_c P(\cdot)=1$,
    \begin{eqnarray}
        1 \ \leq \ \sum_c \frac{e^{N\phi_q(c,\theta,N)}}{\mathcal{Z}_{q,N}(\theta)}
          \ \leq \ (N+1)\frac{e^{N\phi_q(c_{\max},\theta,N)}}{\mathcal{Z}_{q,N}(\theta)}\, .
    \end{eqnarray}
    Taking logarithms and dividing by $N$,
    \begin{eqnarray}
        \frac{\ln\mathcal{Z}_{q,N}(\theta)}{N} \ \leq \ \frac{\ln(N+1)}{N} \ + \ \phi_q(c_{\max},\theta,N)\, .
\label{eq:appendix_Z_upper_bound}
    \end{eqnarray}
\end{itemize}
Since $\ln(N+1)/N \to 0$, squeezing~\eqref{eq:appendix_Z_lower_bound} and~\eqref{eq:appendix_Z_upper_bound} yields\footnote{Strictly, the discrete maximum over $c\in\{0,1/N,\ldots,1\}$ differs from the continuous maximum over $[0,1]$ by $\mathcal{O}(1/N)$; see Ref.~\cite{Dembo2010}.}
\begin{eqnarray}
    \lim_{N\to\infty}\frac{\ln\mathcal{Z}_{q,N}(\theta)}{N}
    \ = \ \lim_{N\to\infty}\max_{c\in[0,1]}\phi_q(c,\theta,N)\, ,
    \label{eq:appendix_free_entropy_limit}
\end{eqnarray}
which is Eq.~\eqref{eq:free_entropy_limit} of the main text.

\subsection*{Step 3: Large-deviation form of $P$}

Combining the probability bounds~\eqref{eq:appendix_prob_bounds} with the partition function bounds~\eqref{eq:appendix_Z_lower_bound}--\eqref{eq:appendix_Z_upper_bound} yields
\begin{eqnarray}
    -\frac{2\ln(N+1)}{N} \ + \ \phi_q(c,\theta,N) \ - \ \phi_q(c_{\max},\theta,N) 
    &\leq&  \frac{1}{N}\ln P\!\left(\frac{L(G)}{N}=c\right) \nonumber\\[1mm]
    &\leq&  \phi_q(c,\theta,N) \ - \  \phi_q(c_{\max},\theta,N) \ + \  \frac{\ln(N+1)}{N}\, .
    \label{appendix_IV.50.mn}
\end{eqnarray}

For fixed $\theta$ and $q$, the second term in~\eqref{eq:appendix_effective_free_entropy} satisfies $\frac{1}{N}\ln(1+(1-q)\theta Nc)\to 0$ as $N\to\infty$, so $\phi_q(c,\theta,N)\to H_b(c)$ and $c_{\max}\to 1/2$. Taking $N\to\infty$ in~\eqref{appendix_IV.50.mn},
\begin{eqnarray}
    \lim_{N\to\infty}\frac{1}{N}\ln P\!\left(\frac{L(G)}{N}=c\right)
    \ = \ H_b(c) \ - \ H_b(1/2)\, ,
\end{eqnarray}
or equivalently
\begin{equation}
    P\!\left(\frac{L(G)}{N}=c\right)
    \ \asymp \ e^{-N\left[H_b(1/2)-H_b(c)\right]}\, ,
    \label{B.12.fg}
\end{equation}
where $a_N \asymp b_N$ means that $\frac{1}{N}(\ln a_N - \ln b_N) \to 0$. 
This establishes Eq.~\eqref{eq:large_deviation_result} of the main text, with rate function $I(c) = H_b(1/2) - H_b(c) \geq 0$. 
Relation~(\ref{B.12.fg}) represents the asymptotic equipartition property~\cite{Cover2006,Jizba2024a} associated with the Erd\H{o}s--R\'enyi model.

\section{The average number of links}
\label{sec:average_links}
The average number of links is computed as follows
\begin{equation}
    \langle L \rangle \ = \  \frac{\sum_{k=0}^{N}k\binom{N}{k}\exp_{2-q}(-\theta k)}{\mathcal{Z}_q^N(\theta)}\, .
\end{equation}
Next, we can use the identity $k\binom{N}{k} = N\binom{N-1}{k-1}$, and the fact that we can write the sum from $k=1$ as the term with $k=0$ does not contribute. The numerator then is

\begin{eqnarray}
    \sum_{k=0}^{N}k\binom{N}{k}\exp_{2-q}(-\theta k) &=& N\sum_{k=1}^N \binom{N-1}{k-1} (1-(q-1)\theta k)^{1/(q-1)} \nonumber \\[2mm] &=& N\sum_{k=0}^{N-1} \binom{N-1}{k} (1-(q-1)\theta (k+1))^{1/(q-1)} \nonumber \\[2mm] &=&
    N\sum_{k=0}^{N-1} \binom{N-1}{k} (1-(q-1)\theta - (q-1)\theta k)^{1/(q-1)} \nonumber \\[2mm] 
    &=&
    N(1 - (q-1)\theta)^{1/(q-1)} \sum_{k=0}^{N-1} \binom{N-1}{k} \left(1 - (q-1)\frac{\theta}{1-(q-1)\theta}k\right)^{1/(q-1)} \nonumber \\[2mm]
    &=&
    N\exp_{2-q}\left(-\theta\right) \mathcal{Z}_q^{N-1}\left(\frac{\theta}{1-(q-1)\theta}\right).
\end{eqnarray}

The average number of links can be then written as
\begin{equation}
    \langle L \rangle \ = \   N\exp_{2-q}\left(-\theta\right)  \frac{\mathcal{Z}_q^{N-1}\left(\frac{\theta}{1-(q-1)\theta}\right)}{\mathcal{Z}_q^N(\theta)}
    \label{eq:avg_num_links_tsallis}\, .
\end{equation}

\section{Direct computation of $\langle a_{ij} \rangle$}
\label{sec:direct_computation_of_avg_link_density}
%
\begin{eqnarray}
        \langle a_{rs} \rangle &=& \frac{1}{\mathcal{Z}_{q,N}(\theta)}\sum_{\{a_{ij}\}} a_{rs} \exp_{2-q}\Big(-\theta \sum_{i=1}^n\sum_{j < i} a_{ij}\Big) \ = \ \frac{1}{\mathcal{Z}_{q,N}(\theta)}\sum_{a_{rs}=0,1}a_{rs} \sum_{\substack{\{a_{ij}\}_{i<j}\\(i,j)\neq (r,s)}} \exp_{2-q}\Big(-\theta \sum_{i=1}^n\sum_{j < i} a_{ij}\Big) \nonumber \\[2mm]
        &=& \frac{1}{\mathcal{Z}_{q,N}(\theta)}\sum_{\substack{\{a_{ij}\}_{i<j}\\(i,j)\neq (r,s)}}\exp_{2-q}\Big(-\theta -\theta \sum_{\substack{i<j\\(i,j)\neq (r,s)}} a_{ij}\Big) \nonumber \\[2mm] 
        &{=}& \frac{1}{\mathcal{Z}_{q,N}(\theta)}\exp_{2-q}(-\theta)\sum_{\substack{\{a_{ij}\}_{i<j}\\(i,j)\neq (r,s)}}\exp_{2-q}\Big(-\frac{\theta}{1-(q-1)\theta} \sum_{\substack{i<j\\(i,j)\neq (r,s)}} a_{ij}\Big)\nonumber  \\[2mm]
        &=& \frac{1}{\mathcal{Z}_{q,N}(\theta)}\exp_{2-q}(-\theta)\mathcal{Z}_{q,N-1}\left(\frac{\theta}{1-(q-1)\theta}\right),
    \label{eq:avg_links_directly}
\end{eqnarray}
where on the third line, we utilized formula \eqref{eq:q_exp_sum_formula}, and on the last line we used the definition of~$\mathcal{Z}_{q,N}(x)$ [see Eq.~(\ref{IV.31.kl})].
%

\section{Computation of higher moments $\langle a_{i_1 j_1} \dots a_{i_k j_k} \rangle$}
\label{sec:higher_moments}
%
Similarly to how we computed $\langle a_{ij} \rangle$ in \eqref{eq:avg_links_directly}, we can compute higher moments. First, let us realize, that $a_{ij} = a_{ij}^2 = a_{ij}^n$ for any $n\in \mathbb{N}$, since we are dealing with binary variables. Therefore, it suffices to consider distinct links.

Let us have $k$ distinct links $i_1 < j_1, \dots, i_k < j_k$. Then we can write
\begin{eqnarray}
    \langle a_{i_1 j_1} \dots a_{i_k j_k} \rangle &=& \frac{1}{\mathcal{Z}_{q,N}(\theta)}\sum_{\{a_{ij}\}} a_{i_1 j_1} \dots a_{i_k j_k} \exp_{2-q}\left(-\theta \sum_{i<j} a_{ij}\right) \nonumber \\[2mm] &=& \sum_{a_{i_1 j_1}=0,1} \dots \sum_{a_{i_k j_k}=0,1} a_{i_1 j_1} \dots a_{i_k j_k} \sum_{\substack{\{a_{ij}\}_{i<j}\\(i,j)\notin \{(i_\alpha,j_\alpha)\}}} \exp_{2-q}\left(-\theta \sum_{i<j} a_{ij}\right).
\end{eqnarray}
Now, the only summands contributing to the sum are those for which $a_{i_1 j_1} = \dots = a_{i_k j_k} = 1$. Therefore, we can rewrite the sum as
\begin{eqnarray}
    \langle a_{i_1 j_1} \dots a_{i_k j_k} \rangle &=& \frac{1}{\mathcal{Z}_{q,N}(\theta)}\sum_{\substack{\{a_{ij}\}_{i<j}\\(i,j)\notin \{(i_\alpha,j_\alpha)\}}} \exp_{2-q}\left(-k\theta -\theta \sum_{\substack{i<j\\(i,j)\notin \{(i_\alpha,j_\alpha)\}}} a_{ij}\right) \nonumber \\[2mm]
    &{=}& \frac{1}{\mathcal{Z}_{q,N}(\theta)}\exp_{2-q}\left(-k\theta\right)\sum_{\substack{\{a_{ij}\}_{i<j}\\(i,j)\notin \{(i_\alpha,j_\alpha)\}}} \exp_{2-q}\left(-\frac{\theta}{1-k(q-1)\theta} \sum_{\substack{i<j\\(i,j)\notin \{(i_\alpha,j_\alpha)\}}} a_{ij}\right) \nonumber \\[2mm]
    &=& \frac{1}{\mathcal{Z}_{q,N}(\theta)}\exp_{2-q}\left(-k\theta\right)\mathcal{Z}_{q,N-k}\left(\frac{\theta}{1-k(q-1)\theta}\right),
\end{eqnarray}
where on the second line  we used the identity~\eqref{eq:q_exp_sum_formula}.

\section{The $q \approx 1$ expansion of $\langle a_{ij} \rangle$ and $\mbox{Cov}(a_{ij}, a_{kl})$}
\label{sec:q_approx_1_num_links}

As discussed, the first and second joint cumulants can be calculated as 
\begin{eqnarray}
    \frac{\partial \ln \mathcal{Z}_{q,N}}{\partial h_{kl}}\bigg|_{\{h_{ij}\}=0} &=& \langle a_{kl}\rangle\,,  \\[2mm]
    \frac{\partial^2 \ln \mathcal{Z}_{q,N}}{\partial h_{kl} \partial h_{mn}}\bigg|_{\{h_{ij}\}=0} &=& \langle a_{kl} a_{mn} \rangle - \langle a_{kl} \rangle \langle a_{mn} \rangle \nonumber\\[1mm]
    &=& \mbox{Cov}(a_{kl}, a_{mn})\, ,
\end{eqnarray}
where
\begin{eqnarray}
    \ln \mathcal{Z}_{q,N}(\theta, \{h_{ij}\}) \ = \ \ln \mathcal{Z}_{1,N}(\theta, \{h_{ij}\}) \ - \ \frac{1}{2}(q-1)\theta^2 \langle L^2 \rangle_{\text{Sh}} \ + \  \mathcal{O}\!\left((q-1)^2\right)\, .
    \label{eq:ln_partition_fn_q_appendix}
\end{eqnarray}
Since $\mathcal{Z}_{1,N}(\theta, \{h_{ij}\})$ is the Shannonian partition function, we know that
\begin{equation}
    \frac{\partial \ln \mathcal{Z}_{1,N}}{\partial h_{kl}}\bigg|_{\{h_{ij}\}=0} \ = \  \langle a_{kl} \rangle_{\text{Sh}}\, , \qquad
    \frac{\partial^2 \ln \mathcal{Z}_{1,N}}{\partial h_{kl} \partial h_{mn}}\bigg|_{\{h_{ij}\}=0} \ = \  0\,.
    \label{eq:derivatives_shannonian_partition_fn_appendix}
\end{equation}
and also 
\begin{equation}
    \langle a_{ij} \rangle_{\text{Sh}} = \frac{1}{1+e^{\theta - h_{ij}}}\,, \qquad \frac{\partial \langle a_{ij} \rangle_{\text{Sh}}}{\partial h_{ij}} \ = \  \frac{e^{\theta - h_{ij}}}{(1+e^{\theta - h_{ij}})^2} \ = \ \langle a_{ij} \rangle_{\text{Sh}}\left[1-\langle a_{ij} \rangle_{\text{Sh}}\right]\, .
    \label{eq:shannonian_link_probability_appendix}
\end{equation}
We can expand $\langle L^2 \rangle_{\text{Sh}}$ as
\begin{eqnarray}
    \langle L^2 \rangle_{\text{Sh}}\  &=&  \ \sum_{i<j} \langle a_{ij}^2 \rangle_{\text{Sh}} \ + \  \sum_{i<j}\sum_{\substack{k<l \\ (k,l)\neq (i,j)}} \langle a_{ij} a_{kl} \rangle_{\text{Sh}} \nonumber \\[2mm]&=& \ \sum_{i<j} \langle a_{ij} \rangle_{\text{Sh}} \ + \  \sum_{i<j}\sum_{\substack{k<l \\ (k,l)\neq (i,j)}} \langle a_{ij} \rangle_{\text{Sh}} \langle a_{kl} \rangle_{\text{Sh}}\, .
\end{eqnarray}
The first derivative with respect to $h_{mn}$, evaluated at $\{h_{ij}\} = 0$, is
\begin{eqnarray}
    \frac{\partial\langle L^2 \rangle_{\text{Sh}}}{\partial h_{mn}}\bigg|_{\{h_{ij}\}=0} \ &=& \ \left.\left[\frac{\partial\langle a_{mn} \rangle_{\text{Sh}}}{\partial h_{mn}} \ + \  2\sum_{\substack{k<l \\ k\neq m,\,l\neq n}}\langle a_{kl} \rangle_{\text{Sh}} \frac{\partial\langle a_{mn} \rangle_{\text{Sh}}}{\partial h_{mn}}\right]\right|_{\{h_{ij}\}=0}\nonumber\\[2mm]
    &=& \frac{e^\theta}{\left(1+e^\theta\right)^2} \ + \  2(N-1)\frac{1}{1+e^\theta}\frac{e^\theta}{\left(1+e^\theta\right)^2} \ = \  \frac{e^\theta}{\left(1+e^\theta\right)^3}\left[2N \ - \  1 \ + \  e^\theta\right]\, .
\end{eqnarray}
Combining Eqs.~\eqref{eq:ln_partition_fn_q_appendix}, \eqref{eq:derivatives_shannonian_partition_fn_appendix}, and \eqref{eq:shannonian_link_probability_appendix}, we obtain the final expression in Eq.~\eqref{eq:er_marginal_link_probability}.

For the covariance between distinct links $(m,n)\neq(p,q)$, we compute
\begin{eqnarray}
    \frac{\partial^2\langle L^2 \rangle_{\text{Sh}}}{\partial h_{mn} \partial h_{pq}}\bigg|_{\{h_{ij}\}=0} \ = \ 2\,\frac{\partial \langle a_{mn} \rangle_{\text{Sh}}}{\partial h_{pq}}\bigg|_{\{h_{ij}\}=0}\frac{\partial \langle a_{pq} \rangle_{\text{Sh}}}{\partial h_{mn}}\bigg|_{\{h_{ij}\}=0} \ = \ 2\frac{e^{2\theta}}{\left(1+e^\theta\right)^4}\,  .
\end{eqnarray}
This implies
\begin{eqnarray}
    \mbox{Cov}(a_{mn}, a_{pq}) \ &=& \  (1-q)\frac{\theta^2 e^{2\theta}}{\left(1+e^\theta\right)^4}\, .
\end{eqnarray}

\end{widetext}

\section{The phase transition condition}
\label{sec:phase_transition_condition}
\begin{figure}[ht]
    \centering
    \includegraphics[width=0.47\textwidth]{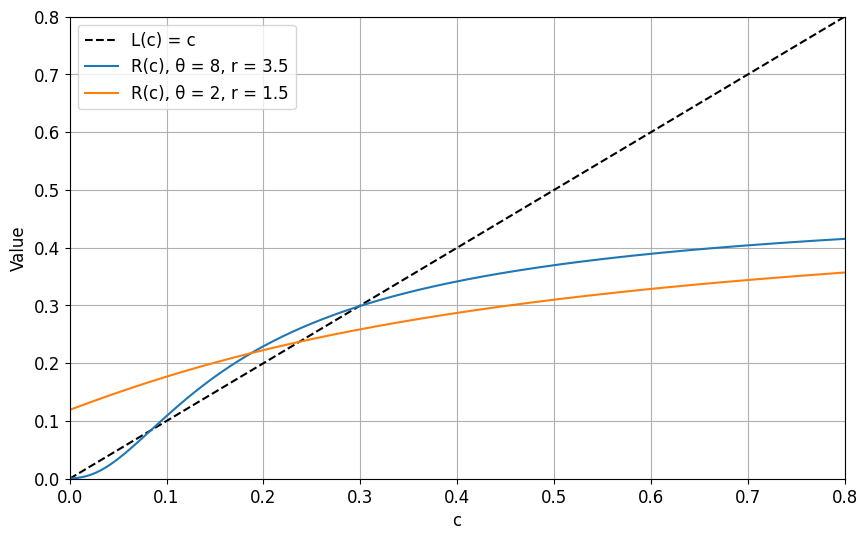}
    \caption{The left-hand side $L(c) = c$ and the right-hand side $R(c)$ of the saddle-point equation~\eqref{eq:saddle_point_equation} for two distinct combinations of $r$ and $\theta$. To have more than one intersection (solutions to the saddle-point equation), the necessary condition is that the derivative of $R(c)$ must be larger than 1 on some interval.}
    \label{fig:saddle_point}
\end{figure}

%
To gain further insight into the phase transition described in Sec.~\ref{sec:phase_transition}, we examine the saddle-point equation~\eqref{eq:saddle_point_equation} in more detail. By rewriting the right-hand side in the form
\begin{equation}
    R(c) \ = \ \left[1 \ + \  \exp\left(\frac{1}{\frac{1}{\theta}+r c}\right)\right]^{-1},
    \label{eq:saddle_point_equation_rhs}
\end{equation}
we note that $R(c)$ is bounded between $0$ and $1$ and is discontinuous at $c = -1/(r\theta)$, which is negative since both $r$ and $\theta$ are positive. It is straightforward to see that on the interval $(-1/(r\theta), +\infty)$, $R(c)$ is a differentiable, monotonically increasing function of $c$ that approaches $1/2$ as $c \to +\infty$. Consequently, the only way for $R(c)$ to intersect the left-hand side $L(c) = c$ more than once is if the derivative of $R(c)$ with respect to $c$ exceeds 1 (the derivative of $L(c)$) over some interval. If this condition is met, there exists a value of $\theta$ for which $R(c)$ has multiple intersections with $L(c)$. An illustration is provided in Fig.~\ref{fig:saddle_point}.

To assess this condition more quantitatively, we note that in Eq.~\eqref{eq:saddle_point_equation_rhs}, $\theta$ acts as a shift along the $c$-axis and thus does not alter the shape of $R(c)$. Consequently, the above condition is determined solely by the value of $r$, and it is sufficient to examine the criterion
\begin{equation}
    \frac{\partial}{\partial c} \left[1  \ + \ \exp\left(\frac{1}{r c}\right)\right]^{-1} \ > \  1\, .
    \label{eq:phase_transition_condition}
\end{equation}
Let us denote $f(c,r) = \left[1 + \exp\left(\frac{1}{r c}\right)\right]^{-1}$ and recognize that it can be written using a {\em logistic function} $\sigma(x) = \frac{1}{1+e^{-x}}$ as $f(c,r) = \sigma\left(-\frac{1}{r c}\right)$. In passing we note that the logistic function satisfies the following useful identities:
\begin{eqnarray}
    &&\sigma(-x) \ = \ 1 \ - \  \sigma(x)\, , \label{eq:logistic_shift}  \\[2mm]
    &&\frac{d}{dx}\sigma(x) \ = \  \sigma(x)[1-\sigma(x)]\, , \label{eq:logistic_derivative}  \\[2mm]
    &&2\sigma(x) \ = \  1 \ + \  \tanh\left(\frac{x}{2}\right)\, , \label{eq:logistic_tanh}  \\[2mm]
    &&\tanh\left(\frac{x}{2}\right) \ = \ \sigma(x) \ - \  \sigma(-x)\, . \label{eq:logistic_tanh_diff}
\end{eqnarray}
Using \eqref{eq:logistic_shift} and \eqref{eq:logistic_derivative}, we rewrite the condition~\eqref{eq:phase_transition_condition} as
\begin{eqnarray}
    \frac{\partial}{\partial c} f(c,r) &=&  \frac{\partial}{\partial c} \sigma\left(-\frac{1}{r c}\right)\nonumber \\[2mm] 
    &=&  \sigma\left(-\frac{1}{r c}\right)\sigma\left(\frac{1}{r c}\right) \frac{1}{r c^2} \ > \  1\, .
    \label{eq:phase_transition_condition_appendix_2}
\end{eqnarray}
We recognize that this expression is odd in $c$, and therefore it is enough to check the condition for $c>0$. Since we consider $r>0$, we have $\frac{\partial}{\partial c} f(c,r) > 0$ and it is easy to check that 
\begin{eqnarray}
\lim_{c\to 0^+} \frac{\partial}{\partial c} f(c,r) \ = \  \lim_{c\to +\infty} \frac{\partial}{\partial c} f(c,r) \ = \ 0\, .
\end{eqnarray}
Since $\frac{\partial}{\partial c} f(c,r)$ is differentiable (and hence also continuous), it suffices to check the condition at its maximum. 

The condition for the maximum of $\frac{\partial}{\partial c} f(c,r)$ is given by
\begin{equation}
    \frac{\partial^2}{\partial c^2} f(c,r) \ = \  0\, .
\end{equation}
By evaluating the derivative, we obtain
\begin{equation}
    \sigma\left(\frac{1}{rc}\right) \ - \  \sigma\left(-\frac{1}{rc}\right) \  - \  2rc \ = \  0\, ,
\end{equation}
and with the help of the identity~\eqref{eq:logistic_tanh_diff}
\begin{equation}
    \tanh\left(\frac{1}{2rc}\right) \ = \ 2rc\, .
    \label{eq:maximum_condition}
\end{equation}
The maximum of $\frac{\partial}{\partial c} f(c,r)$ is therefore attained at the value of $c$ satisfying~Eq.~\eqref{eq:maximum_condition}. 
The condition~\eqref{eq:phase_transition_condition_appendix_2} can then be rewritten as
\begin{equation}
    r \ > \ \frac{(2rc)^2}{4\sigma\left(-\frac{1}{rc}\right)\sigma\left(\frac{1}{rc}\right)}\, ,
\end{equation}
which can be further rewritten using the identity \eqref{eq:logistic_tanh} and the fact that $\tanh(x)$ is an odd function as
\begin{equation}
    r \ > \  \frac{(2rc)^2}{1-\tanh^2\left(\frac{1}{2rc}\right)}\, .
\label{eq:phase_transition_condition_appendix_3}
\end{equation}
Combining Eqs.~\eqref{eq:maximum_condition} and~\eqref{eq:phase_transition_condition_appendix_3}, we see that it suffices to identify a solution $x^*$ of 
$\tanh(1/x^*) = x^*$. 
Condition~\eqref{eq:phase_transition_condition_appendix_3} then implies 
$r > \frac{(x^*)^2}{1-(x^*)^2}$, 
which completes the proof.

\section{The critical point}
\label{sec:critical_point}
%
The second derivative of $\phi_r(c, \theta)$ is given by
\begin{equation}
    \frac{\partial^2}{\partial c^2} \ \! \phi_r(c, \theta) \ = \  \frac{1}{c(c-1)} \ + \  \frac{r}{\frac{1}{\theta} + rc}\, .
    \label{eq:saddle_point_equation_second_derivative}
\end{equation}
Setting to zero and rearranging, we get
\begin{equation}
    1 \ + \  cr\theta(2-\theta) \ - \ c^2r\theta^2(r+1) \ = \  0\, .
\end{equation}
We want to find the smallest $\theta$, for which this equation has a solution. That is equivalent to finding such $\theta$, for which the discriminant of the quadratic equation is zero. This yields
\begin{equation}
    r^2\theta^2(2-\theta)^2 \ - \  4r\theta^2(r+1) \ = \  0\, .
\end{equation}
Assuming $\theta\neq0$ and using the $r$ value of the critical point $r_{cp}$, we find
\begin{equation}
    \theta_{cp} \ = \  2 \ + \  \sqrt{\frac{4(r_{cp}+1)}{r_{cp}}}\, .
\end{equation}

\section{$q$-ER model -- Percolation}
\label{sec:qERM_clustering_percolation}

\begin{figure}[t]
    \centering
    \includegraphics[width=0.48\textwidth]{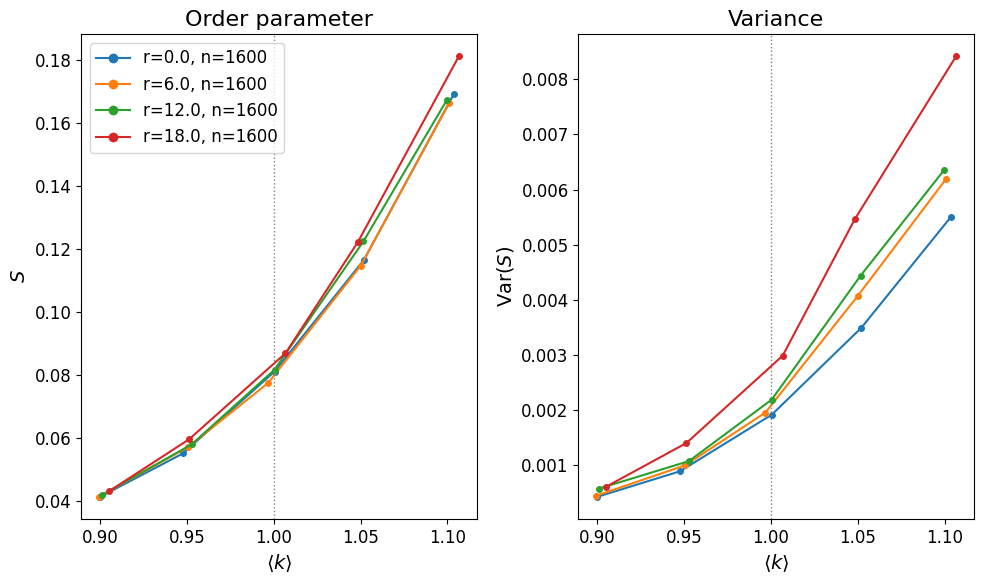}
    \caption{Order parameter $S$ (left) and its variance $\mathrm{Var}(S)$ (right) as a function of target average degree $k^*$ near the percolation threshold, for $n=1600$ nodes and $r \in \{0, 6, 12, 18\}$. The mean size of the giant component is insensitive to $r$, while the variance grows substantially with increasing $r$, indicating enhanced critical fluctuations.}
    \label{fig:er_percolation}
\end{figure}

For the percolation study, we generated ensembles of 1000 graphs with $n = 1600$ nodes for target average degrees $k^* \in \{0.9, 0.95, 1.0, 1.05, 1.1\}$ (spanning the percolation threshold at $k^* = 1$) and $r \in \{0, 6, 12, 18\}$. We tracked the order parameter $S$ (fraction of nodes in the largest connected component) and its variance $\mathrm{Var}(S)$. As shown in Fig.~\ref{fig:er_percolation}, the mean value of $S$ is virtually identical across all values of $r$, confirming that the large-$N$ behavior of the giant component is the same as in the Shannon case. However, the variance of $S$ grows monotonically with $r$: near the percolation threshold, ensembles with larger $r$ exhibit significantly larger fluctuations in the size of the largest connected component. This suggests that while the $q$-ER model does not shift the percolation threshold, it does amplify the critical fluctuations, consistent with the enhanced link correlations identified analytically in Eq.~\eqref{eq:er_edge_covariance}.

\section{$q$-Exponential Configuration Model -- Empirical Results}
\label{sec:conf_model_empirical_results}

As described in Sec.~\ref{sec:tsallis_park_newman_network_properties}, we study the $q$-exponential CM for target average node degrees $k^* \in \{1, 3, 5, 7, 9\}$ and a range of values of $r = (1-q)N$ from $r=-30$ up to $r=9$ (for $r>9$, we often start hitting the irreconstructible region). For each combination of $k^*$ and $r$, we determine the additive shift $\theta_{\mathrm{shift}}$ by bisection so that the sampled ensemble reproduces the target density, and then generate 1000 networks with $n=1000$ nodes using the Metropolis--Hastings algorithm.

\subsection{Fitting the shift parameter}

\begin{figure}[t]
    \centering
    \includegraphics[width=0.48\textwidth]{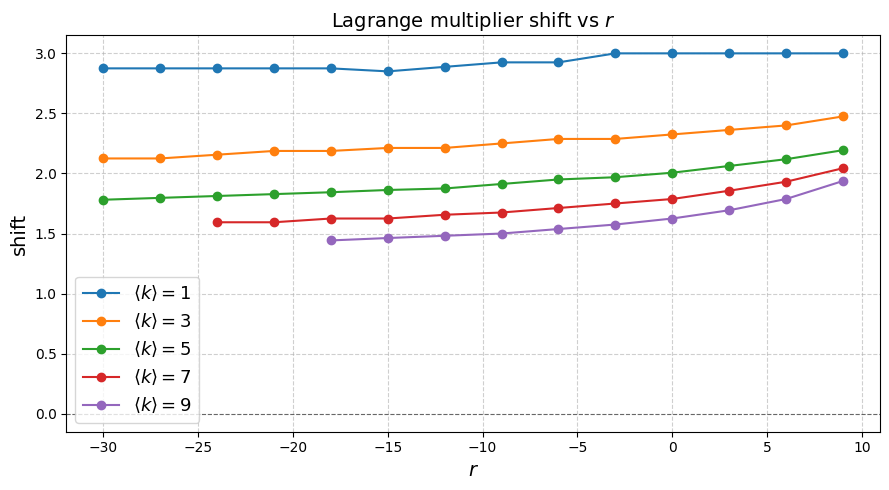}
    \caption{Lagrange multiplier shift $\theta_{\mathrm{shift}}$ as a function of $r$ for target average degrees $k^* \in \{1, 3, 5, 7, 9\}$. The shift increases monotonically with $r$ for all target degrees, reflecting that larger $|r|$ requires a stronger compensation to maintain the prescribed density.}
    \label{fig:conf_model_shifts_vs_r}
\end{figure}

Before comparing structural properties, it is instructive to examine how the required Lagrange multiplier shift $\theta_{\mathrm{shift}}$ depends on $r$. As shown in Fig.~\ref{fig:conf_model_shifts_vs_r}, the shift increases monotonically with $r$ for all target average degrees $k^* \in \{1, 3, 5, 7, 9\}$: moving from negative to positive $r$ (i.e.\ from $q > 1$ toward $q < 1$) requires a progressively larger additive shift to maintain the target density. Networks with lower target density (note that higher $k^*$ corresponds to a denser graph, so curves for lower $k^*$ sit higher) require a larger absolute shift overall. 

\subsection{Degree distributions}

An important aspect of our experimental setup is that the parameters $\theta_i$ are \emph{not} fitted individually for each node and each value of $r$. Instead, a single set of $\theta_i$ values is sampled from the normal distribution (with $\sigma = 2.28$, $\mu = \sigma^2/2$) and then kept fixed across all values of $r$; only the common additive shift $\theta_{\mathrm{shift}}$ is adjusted to recover the target average degree. As a consequence, the degree sequence is not enforced exactly: different values of $r$ induce different link probabilities from the same $\theta_i$ profile, leading to differences in the resulting degree distributions.

\begin{figure*}[t]
    \centering
    \includegraphics[width=0.96\textwidth]{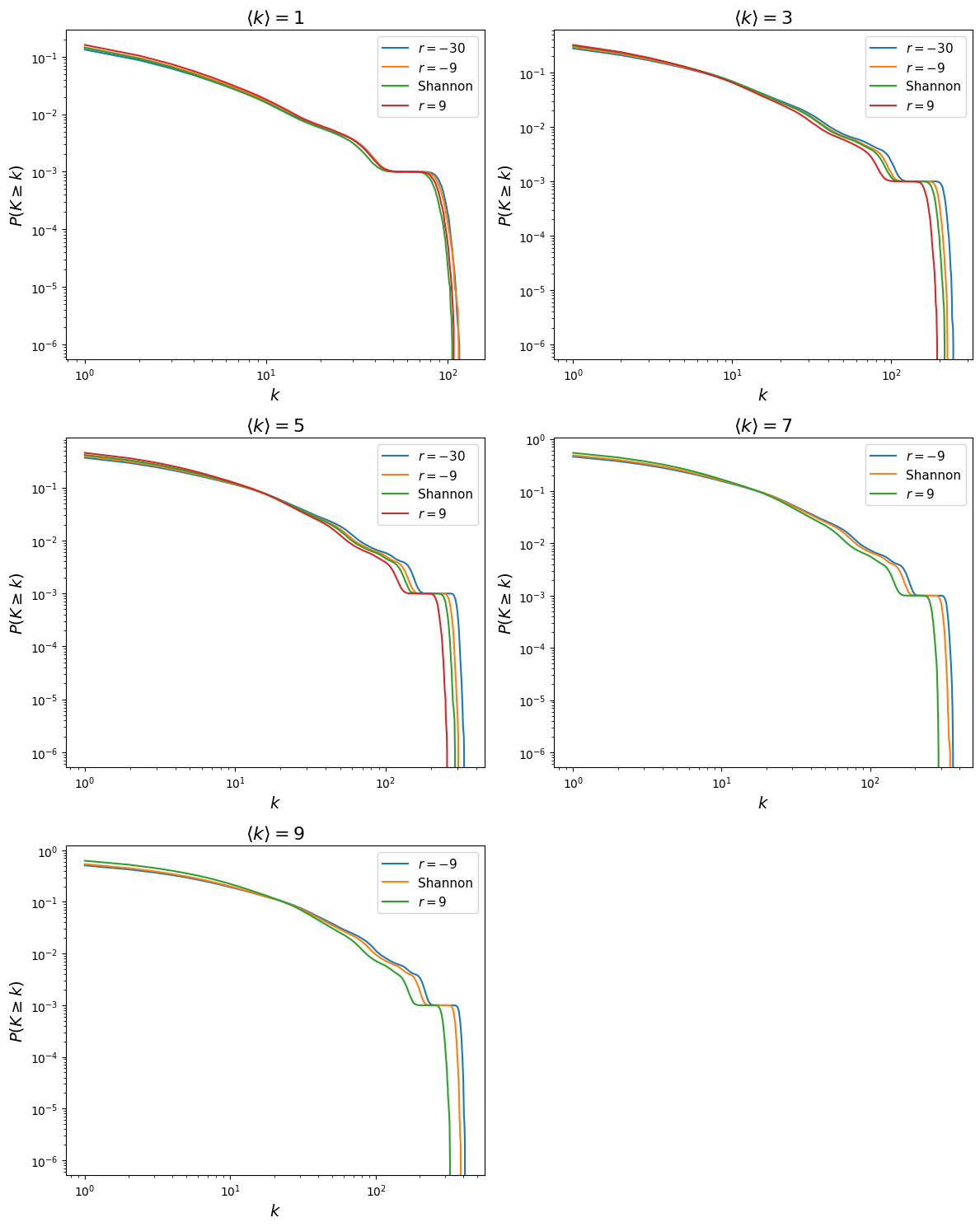}
    \caption{Complementary cumulative degree distributions for the $q$-exponential CM at target average degrees $k^* \in \{1, 3, 5, 7, 9\}$ and selected values of $r \in \{-30, -9, \text{Shannon}, 9\}$. Note that the parameters $\theta_i$ are sampled once and held fixed across all $r$; only the common shift $\theta_{\mathrm{shift}}$ is tuned to match the target average degree. Distributions for different $r$ values are nearly identical at low target degrees but begin to diverge in the tail as the target degree increases, reflecting the growing influence of the non-Shannon coupling on the individual link probabilities.}
    \label{fig:conf_model_degree_dist}
\end{figure*}

Fig.~\ref{fig:conf_model_degree_dist} shows the complementary cumulative degree distributions for all five target average degrees and a representative subset of $r$ values. For low target degrees ($k^* = 1$), the distributions for different values of $r$ remain nearly indistinguishable. As the target degree increases, however, systematic differences between curves for different $r$ values emerge, particularly in the tail of the distribution. This is a direct consequence of the shared-$\theta_i$ setup: at higher densities the non-Shannon coupling has a stronger effect on the individual link probabilities, so the resulting degree sequence deviates more noticeably from the Shannonian baseline as $r$ varies.

\subsection{Tunable clustering}

\begin{figure}[H]
    \centering
    \includegraphics[width=0.48\textwidth]{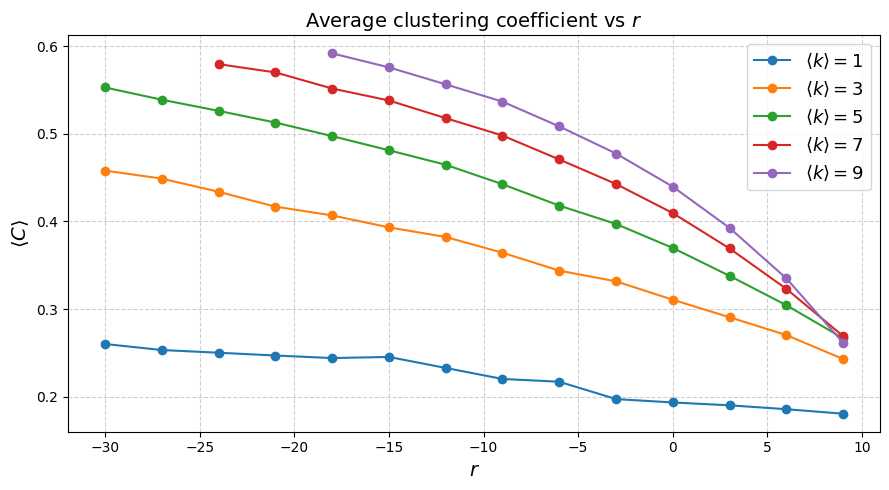}
    \caption{Average clustering coefficient $\langle C \rangle$ as a function of $r$ for target average degrees $k^* \in \{1, 3, 5, 7, 9\}$. The Shannon ($r=0$) result is marked for reference. Clustering decreases monotonically with increasing $r$, and the accessible range grows considerably with $k^*$, demonstrating that $r$ independently controls the clustering at fixed target density.}
    \label{fig:conf_model_clustering_vs_r}
\end{figure}

Fig.~\ref{fig:conf_model_clustering_vs_r} displays the average clustering coefficient $\langle C \rangle$ as a function of $r$ for each target average degree. A clear and consistent trend is visible across all values of $k^*$: increasing $r$ (i.e.\ moving from $q > 1$ toward $q < 1$) leads to a monotonic decrease in average clustering. The range of attainable clustering values widens substantially with increasing $k^*$, corroborating the result of Sec.~\ref{sec:tsallis_park_newman_network_properties} that the parameter $r$ acts as an independent knob for clustering at fixed degree sequence. This tunability is entirely absent in the standard Shannonian CM, where the degree sequence alone determines the clustering.

\subsection{Degree-resolved network properties}

\begin{figure*}[t]
    \centering
    \includegraphics[width=0.9\textwidth]{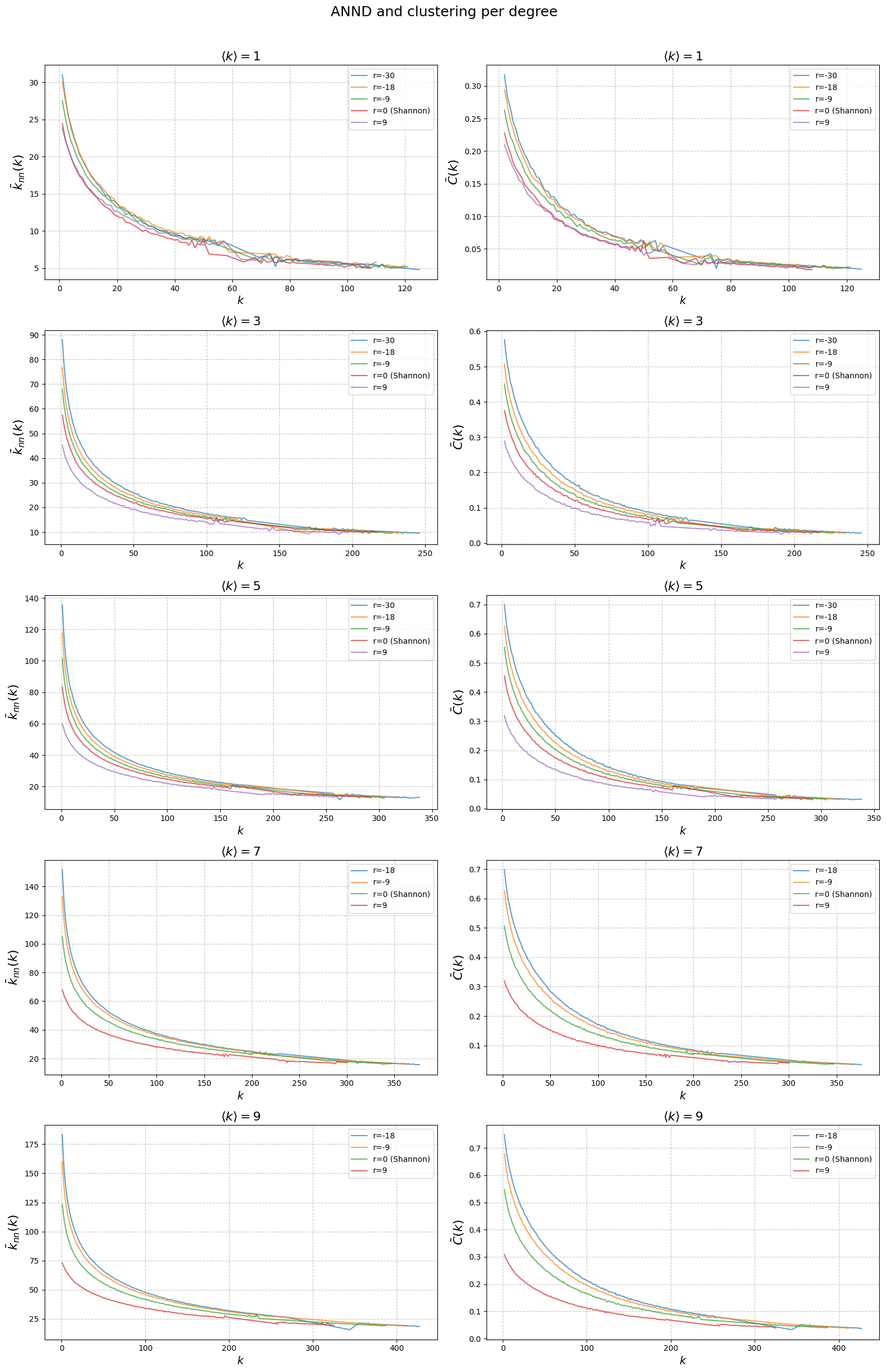}
    \caption{Average nearest-neighbor degree $\bar{k}_{\mathrm{nn}}(k)$ (left column) and local clustering coefficient $\bar{C}(k)$ (right column) as functions of node degree $k$, for target average degrees $k^* \in \{1, 3, 5, 7, 9\}$ (rows) and selected values of $r$. Both quantities decrease with $k$ in all cases, but their overall level shifts systematically with $r$, demonstrating that the non-Shannon parameter independently controls higher-order structural properties beyond the degree sequence.}
    \label{fig:conf_model_network_properties}
\end{figure*}

Finally, Fig.~\ref{fig:conf_model_network_properties} shows the average nearest-neighbor degree $\bar{k}_{\mathrm{nn}}(k)$ and the local clustering coefficient $\bar{C}(k)$ as functions of node degree $k$, for all five target average degrees and several values of $r$. Both profiles exhibit a systematic shift with varying $r$: networks with more negative $r$ (deeper into the $q > 1$ regime) display higher values of both the ANND and the clustering coefficient across all degree classes compared to networks with positive $r$. The shape of the profiles — both quantities decreasing with $k$ — is consistent with disassortative mixing, but the overall level is controlled by $r$. This confirms that the $q$-exponential CM provides a genuinely richer structural repertoire than the Shannonian CM, allowing fine-grained control over higher-order structural properties while respecting the prescribed degree sequence.

\section{Metropolis--Hastings sampling}
\label{sec:metropolis_hastings}
%
The Metropolis--Hastings algorithm~\cite{Robert1999} is a Markov chain Monte Carlo method designed to generate a Markov chain with a prescribed stationary distribution $P(G)$. Starting from a current graph configuration $G_t$, a candidate graph $G'$ is proposed according to a proposal distribution $Q(G' | G_t)$. The proposed graph is then accepted with probability
\begin{equation}
    \alpha \ = \ \min\left(1, \frac{P(G')}{P(G_t)}\frac{Q(G_t|G')}{Q(G'|G_t)}\right),
\end{equation}
while $G_t$ is retained with probability $1 - \alpha$. A sufficient condition for convergence to the correct stationary distribution is that any graph can be reached from any other in a finite number of proposal steps.

In both the $q$-ER model and the $q$-CM, we use a proposal that flips a uniformly at random chosen link (i.e.\ toggles a non-existent link to present, or an existing link to absent). This proposal is symmetric, $Q(G' | G) = Q(G | G')$, so the acceptance probability reduces to
\begin{equation}
    \alpha \ = \ \min\left(1, \frac{P(G')}{P(G_t)}\right).
\end{equation}
Crucially, only the ratio of graph probabilities enters, so the (intractable) partition function cancels.

Because each update changes only a single link, successive graphs in the chain are strongly correlated. To reduce autocorrelations we apply \emph{thinning}: after every $M$ proposed steps — where $M$ is of the order of the total number of possible links — we append one graph to the ensemble. We generate ensembles of 1000 graphs for each set of model parameters studied in this work.

\begin{widetext}
    \section{Link probability in the $q$-exponential CM}
    \label{sec:avg_link_density_cm}
    %
    We derive Eq.~\eqref{eq:avg_link_density_park_newman} of the main text. Starting from the definition and isolating the sum over $a_{kl}$,
    \begin{eqnarray}
            \langle a_{kl} \rangle \mathcal{Z}_{q}(\theta_1, \dots, \theta_N) &=& \sum_{\{a_{ij}\}} a_{kl} \exp_{2-q}\Big[-\sum_{i=1}^n\sum_{j < i}(\theta_i + \theta_j)a_{ij}\Big] \nonumber \\[2mm]
            &=& \sum_{a_{kl}=0,1}a_{kl}\sum_{\substack{\{a_{ij}\}\\(i,j)\neq (k,l)}} \exp_{2-q}\Big[-(\theta_k + \theta_l)a_{kl} \ - \sum_{\substack{i<j\\(i,j)\neq (k,l)}}(\theta_i + \theta_j) a_{ij}\Big] \nonumber \\[2mm]
            &=& \sum_{\substack{\{a_{ij}\}\\(i,j)\neq (k,l)}} \exp_{2-q}\Big[-(\theta_k + \theta_l) \ - \sum_{\substack{i<j\\(i,j)\neq (k,l)}}(\theta_i + \theta_j) a_{ij}\Big] \nonumber \\[2mm]
            &{=}& \exp_{2-q}\!\left(-(\theta_k + \theta_l)\right)\sum_{\substack{\{a_{ij}\}\\(i,j)\neq (k,l)}} \exp_{2-q}\!\left[- \sum_{\substack{i<j\\(i,j)\neq (k,l)}}\frac{\theta_i + \theta_j}{1-(q-1)(\theta_k + \theta_l)} a_{ij}\right] \nonumber \\[3mm]
            &\equiv& \exp_{2-q}\!\left(-(\theta_k + \theta_l)\right)\tilde{\mathcal{Z}}^{(k,l)}_{q}\!\left(\tilde{\theta}^{(k,l)}_1, \dots, \tilde{\theta}^{(k,l)}_N\right),
    \end{eqnarray}
    where the third-to-fourth line uses the $q$-exponential addition formula~\eqref{eq:q_exp_sum_formula}. Dividing both sides by $\mathcal{Z}_q$ yields Eq.~\eqref{eq:avg_link_density_park_newman}.
    \\
\end{widetext}

\clearpage
\bibliography{ref}

\end{document}